\documentclass[sigconf, screen]{acmart}
\PassOptionsToPackage{table,xcdraw}{xcolor}
\AtBeginDocument{%
  }

\setcopyright{acmlicensed}
\copyrightyear{2026}
\acmYear{2026}
\setcopyright{cc}
\setcctype{by-nc-nd}
\acmConference[CHI '26]{Proceedings of the 2026 CHI Conference on Human Factors in Computing Systems}{April 13--17, 2026}{Barcelona, Spain}
\acmBooktitle{Proceedings of the 2026 CHI Conference on Human Factors in Computing Systems (CHI '26), April 13--17, 2026, Barcelona, Spain}
\acmDOI{10.1145/3772318.3790616}
\acmISBN{979-8-4007-2278-3/2026/04}

\newcommand{\wg}{women's growth~}
\newcommand{\wge}{women's growth}
\usepackage{hyperref}
\usepackage{hyperxmp}
\usepackage{amsmath}
\usepackage{algorithm}

\usepackage{color}
\usepackage{enumitem}
\usepackage{bbding}
\usepackage{listings}
\usepackage{multirow}
\usepackage{colortbl}
\usepackage{graphicx}
\usepackage{tabularx} %? 
\usepackage{makecell} 
\usepackage{array}  
\usepackage{geometry}
\usepackage{CJK}

\newcommand{\camera}[1]{{\leavevmode\color{black} #1}}

\newcommand{\eg}{{e.g.,\ }}

\newcommand{\ie}{{i.e.,\ }}

\AtBeginDocument{
}

\definecolor{tablerowcolor}{rgb}{0.667,0.667,0.667 }
\definecolor{tablerowcolor2}{rgb}{0,0,0}
\definecolor{visual}{HTML}{e8efd9}
\definecolor{motion}{HTML}{fde7d5}
\definecolor{narrative}{HTML}{e2dce9}
\definecolor{audio}{HTML}{d6ebf2}
\definecolor{bluecrayola}{rgb}{0.12,0.46,1.0}

\begin{document}
\begin{CJK*}{UTF8}{gbsn}

\title[]{When LLMs Enter Everyday Feminism on Chinese Social Media: Opportunities and Risks for Women's Empowerment}

\author{Runhua Zhang}
\orcid{0000-0002-0519-5148}
\affiliation{%
  \institution{The Hong Kong University of Science and Technology}
  \city{Hong Kong SAR}
  \country{China}  
}
\email{runhua.zhang@connect.ust.hk}

\author{Ziqi Pan}
\orcid{0000-0002-5562-8685}
\affiliation{%
  \institution{The Hong Kong University of Science and Technology}
  \city{Hong Kong SAR}
  \country{China}
}
\email{zpanar@connect.ust.hk}

\author{Kangyu Yuan}
\orcid{0009-0001-8460-9651}
\affiliation{%
  \institution{The Hong Kong University of Science and Technology}
  \city{Hong Kong SAR}
  \country{China}
}
\email{kyuanaf@connect.ust.hk}

\author{Qiaoyi Chen}
\orcid{0009-0005-3892-860X}
\affiliation{%
  \institution{The Hong Kong University of Science and Technology}
  \city{Hong Kong SAR}
  \country{China}
}
\email{qchench@connect.ust.hk}

\author{Yulin Tian}
\orcid{0009-0001-0401-777X}
\affiliation{%
  \institution{College of Design and Innovation, Tongji University}
  \city{Shanghai}
  \country{China}
}
\email{yulin_t@tongji.edu.cn}

\author{Huamin Qu}
\orcid{0000-0002-3344-9694}
\affiliation{%
  \institution{The Hong Kong University of Science and Technology}
  \city{Hong Kong SAR}
  \country{China}  
}
\email{huamin@cse.ust.hk}

%1
\author{Xiaojuan Ma}
\orcid{0000-0002-9847-7784}
\affiliation{%
  \institution{The Hong Kong University of Science and Technology}
  \city{Hong Kong SAR}
  \country{China}
}
\email{mxj@cse.ust.hk}

\renewcommand{\shortauthors}{Zhang et al.}
%\acmSubmissionID{8888}
\begin{abstract}
Everyday digital feminism refers to the ordinary, often pragmatic ways women articulate lived experiences and cultivate solidarity in online spaces. In China, such practices flourish on RedNote through discussions under hashtags like ``women's growth''. Recently, DeepSeek-generated content has been taken up as a new voice in these conversations. Given widely recognized gender biases in LLMs, this raises critical concerns about how LLMs interact with everyday feminist practices. Through an analysis of 430 RedNote posts, 139 shared DeepSeek responses, and 3211 comments, we found that users predominantly welcomed DeepSeek's advice. Yet feminist critical discourse analysis revealed that these responses primarily encouraged women to self-optimize and pursue achievements within prevailing norms rather than challenge them. By interpreting this case, we discuss the opportunities and risks that LLMs introduce for everyday feminism as a pathway toward women's empowerment, and offer design implications for leveraging LLMs to better support such practices.
\end{abstract} 




\maketitle

\section{Introduction}
Everyday feminism has become an important avenue for women's empowerment~\cite{schuster2017a}. Rather than collective mobilization and activism~\cite{mueller2021, gallagher2019}, the term describes small and individually enacted acts of resistance that unfold in ordinary life~\cite{kelly2015}. Social media platforms have become a central setting for these practices because they make such mundane expressions visible, shareable, and collectively interpretable~\cite{sun2024, pruchniewska2019, huang2025}. Through posting, commenting, and interacting, women exercise agency to articulate frustrations and achievements related to everyday sexism, gendered expectations, and personal growth. According to feminist scholarship, these everyday feminist practices online enable women to speak from lived experience~\cite{batool2022}, form affective connections through recognition and resonance~\cite{kelly2015}, and collectively develop situated knowledge about what it means to live as a woman and what kinds of futures feel possible~\cite{pruchniewska2019}. In this paper, we refer to such practices on social media as \textit{everyday digital feminism}. 

\begin{figure*}[t]
  \centering
  \includegraphics[width=\linewidth]{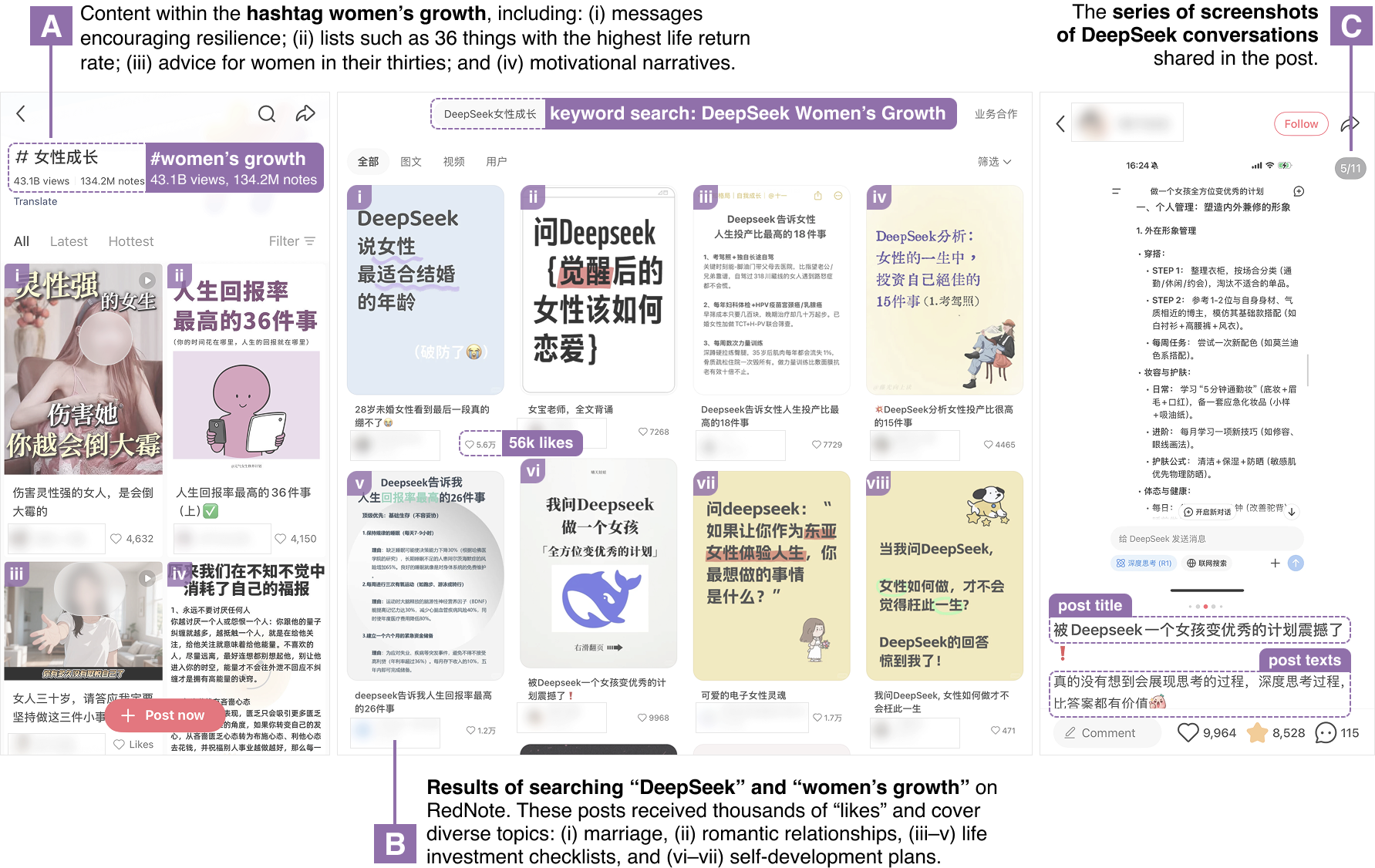}
  \caption{Screenshots of RedNote: (a) Content under the hashtag ``\wge'' as of September 10, 2025, which had accumulated 43.1 billion views and 134.2 million posts. (b) Web-based interface of RedNote, showing diverse search results for ``DeepSeek'' and ``\wge.''  (c) An example post combining ``\wge'' with DeepSeek conversations. Typically, such posts display a cover image followed by a series of screenshots of the conversation with DeepSeek.}
  \label{fig:deepseek_screenshots}
  \Description{Screenshots from RedNote illustrating content under the hashtag ``women's growth'' and search results for ``DeepSeek'' combined with ``women's growth.'' Panel A (left): Shows posts under the hashtag ``women's growth'' with 43.1 billion views and 134.2 million posts. Example content includes: (i) motivational messages about resilience, (ii) popular lists like ``36 things with the highest life return rate,'' (iii) advice for women in their thirties, and (iv) narrative encouragement. Panel B (middle): Search results for ``DeepSeek'' and ``women's growth.'' Posts cover diverse topics such as marriage, romantic relationships, investment checklists, and self-development. Each post has received significant engagement, with ``likes'' ranging from hundreds to tens of thousands. Panel C (right): An example post combining ``women's growth'' with a DeepSeek conversation. The format shows a cover image followed by screenshots of the dialogue with DeepSeek, along with engagement metrics (e.g., likes, shares, comments). The figure highlights the variety of topics, popularity (measured through likes and shares), and integration of AI-based conversations (DeepSeek) within discussions of women's growth on RedNote.}
\end{figure*}

Since everyday feminist discourses frequently focus on women's personal growth and development, their mild and individualized format is often more acceptable within Chinese social media~\cite{sun2024, huang2025, mao2020c, chen2024a}. Among these platforms, RedNote (also known as Xiaohongshu)\footnote{Check Xiaohongshu on Wikipedia: \url{https://en.wikipedia.org/w/index.php?title=Xiaohongshu&oldid=1310407725}}, a popular social media platform with a predominantly young female user base, has become a key site for everyday digital feminism in China~\cite{ngu2025, wang2025}. Hashtags such as ``women's growth'' (\#女性成长) have become focal points for these discourses. Semantically, the term blends a self-development framing, such as becoming more capable or improving oneself, with an affective register that resonates with women's everyday struggles and desires. As of September 10, 2025, this particular hashtag had accumulated more than 43.1 billion views and over 134.2 million posts (see~\autoref{fig:deepseek_screenshots}). Within this vast and ongoing stream of interaction, Chinese women collectively negotiate what ``growth'' and ``development'' mean in their lives, articulating which forms of achievement are valued and which life pathways are imagined as possible or desirable.

Against this backdrop, a new phenomenon has emerged under the hashtag ``women's growth''. Women on RedNote increasingly post screenshots of their conversations with DeepSeek, a widely used Chinese large language model. In these posts, open-ended queries about women's growth, such as life trajectories and development goals, were proposed to DeepSeek. The model's responses to prompts like \textit{Would you choose to get married if you were a woman} or \textit{What can women do to avoid feeling that their life has been in vain}, are then circulated within the community for discussion. As shown in \autoref{fig:deepseek_screenshots}, many of these posts attract substantial engagement (\eg `likes'), indicating their considerable visibility on the platform. 

Given the widely recognized gender biases and growing concerns about the normative orientations embedded in LLMs~\cite{gallegos2024, kotek2023, sachdeva2025}, the entry of LLMs into everyday feminist spaces raises critical concerns. For instance, it remains unclear what kinds of assumptions underlie DeepSeek's responses when offering advice on women's lives. How might LLMs introduce new or reinforce existing interpretations of women's growth and development in specific sociopolitical contexts? What opportunities or risks might their presence bring to everyday feminism? In this paper, we take the phenomenon observed on RedNote as a timely case through which to examine these concerns. Specifically, we propose the following research questions:

\begin{itemize}
    \item \textbf{RQ1}: What \textbf{\textit{topics}} and \textbf{\textit{attitudes}} are reflected in posts that discuss \wg with DeepSeek?  
    \item \textbf{RQ2}: How is \wg framed in DeepSeek responses, in terms of the \textbf{\textit{achievements}} envisioned and the \textbf{\textit{pathways}} suggested? 
    \item \textbf{RQ3}: How does the RedNote community receive and express different \textbf{\textit{stances}} toward DeepSeek responses?
\end{itemize}

To address these questions, we collected relevant RedNote posts, the DeepSeek responses shared, and comments via web scraping. On the one hand, we conducted content analysis of 430 posts (RQ1) and 3211 first-layer comments (RQ3) to examine how DeepSeek were taken up and evaluated. On the other hand, we conducted a feminist critical discourse analysis (FCDA)~\cite{lazar2014} on a set of DeepSeek responses, in order to investigate how women's growth was framed. Our findings show that most posts shared DeepSeek's suggestions for women's self-development plans, covering general life directions to concrete aspirations across different life domains (RQ1, see \autoref{rq1}). Our critical examination of these shared outputs shows that DeepSeek often framed women's growth as a project of self-optimization, focusing on accumulating resources, improving skills, and aligning oneself with prevailing social expectations. At the same time, it paid far less attention to the structural constraints that shape women's choices and opportunities (RQ2, see \autoref{rq2}). Though the community reactions largely legitimized and embraced DeepSeek's suggestions, there were still comments that questioned, negotiated, or resisted DeepSeek's prescriptions (RQ3, see \autoref{rq3}).

Building on these findings, we show that DeepSeek's discourse on women's growth largely aligns with the dominant narratives of self-optimization and self-improvement already prevalent on RedNote~\cite{ge2025, peng2021}, which helps explain why its advice was widely welcomed on the platform. Nevertheless, our discussion also reveals both opportunities and risks that LLMs may bring to this space. We reflect on how LLMs can expand yet also dilute narrative agency grounded in lived experience, how they can facilitate yet also weaken affective solidarity among women, and how they can constrain the production of situated feminist knowledge. Furthermore, we outline design implications for better supporting everyday feminist practices on social media, grounded in a grassroots perspective that is attentive to constrained sociopolitical contexts.

\section{Background and Related Work}
In this section, we first review related work on women's empowerment, including theoretical foundations, the Chinese context, and feminist HCI scholarship, which together shape the orientation of our study. We then turn to everyday digital feminism and gender bias in LLMs, from which we identify our research gaps.

\subsection{Women's Empowerment}
\label{2.1}
\subsubsection{Feminist Conceptualizations of Women's Empowerment} In feminist scholarship, women's empowerment has long been a central concern~\cite{cornwall2016, duflo2012}. Batliwala~\cite{batliwala1993} links empowerment to the transformation of \textit{power} relations, defining it as a process of challenging existing power relations and gaining greater control over the sources of power. Building on this focus on power, Kabeer~\cite{kabeer1999} conceptualizes empowerment in terms of \textit{the ability to make choices}. For Kabeer, empowerment is the process through which those who have been denied the capacity to make strategic life choices come to acquire it. Such ability to make choices is reflected in three interrelated components: \textit{resources}: the material, social, and human conditions that enable people to exercise choice; \textit{agency}: the ability to define one's goals and act upon them; and \textit{achievements}: the outcomes of these choices in terms of well-being and equality. 

By specifying power in terms of women's ability to make choices and by articulating its three components, Kabeer's framework enables the analysis of which choices women are able to make in their lives, and how these choices reflect the shifts in power. Therefore, the framework has been widely used for evaluating or measuring empowerment~\cite{costa2023}. In our study, we leverage Kabeer's framework as the primary theoretical lens. When women turn to LLMs for advice on growth and self-development, the action plans generated by LLMs implicitly or explicitly articulate what resources women should mobilize, how they should exercise agency, and which goals they should pursue. Therefore, Kabeer's framework can offer a structured way to deconstruct and analyze these textual prescriptions (detailed in \autoref{methods}).  

\subsubsection{Women's Empowerment in China}
``women's empowerment'' in the Chinese context is largely shaped by its state-led orientation~\cite{wu2019, liao2021, zheng2010}. It is typically framed in terms of improving women's capabilities such as participation, skills, and self-development in ways that align with national development goals~\cite{zhu2021}. Through such top-down initiatives, many urban women have gained greater access to education, formal employment, and legal protections~\cite{chen_rise_2024}. At the same time, the state's emphasis on social stability and harmony means that large-scale bottom-up feminist activism is uncommon, and organizations advocating structural gender change tend to operate with limited visibility~\cite{li2025a, chen2024a}. 

% also shape how women's empowerment can be understood to ordinary people.
On the other hand, cultural factors such as deeply rooted Confucian norms contribute to public backlash against feminism, and explicitly political feminist claims often become points of contention~\cite{yang2018, qin2024}. Prior research has examined how such controversies unfold online. For example, Qin et al.~\cite{qin2024} showed how gender-based violence is discursively reframed into gender-blind narratives, while Deng et al.~\cite{deng2024b} analyzed tensions between incivility and constructiveness in debates over women's everyday needs, such as calls for more public toilets for women. These dynamics make it difficult to articulate overtly political claims. As a result, feminist discourse in China is typically expressed in gentle and indirect ways rather than through open confrontation~\cite{sun2024, huang2025, mao2020c, chen2024a}. These sociopolitical conditions structure the particular forms that everyday feminism can take in China, as we further discuss in \autoref{2.2}.

\subsubsection{Women's Empowerment in HCI}
With the emergence of third-wave HCI~\cite{harrison2007} and its emphasis on values, agency, and social justice, the concept of women's empowerment has increasingly entered HCI discourse. Drawing on feminist theory, Bardzell's Feminist HCI agenda~\cite{bardzell2010a, bardzell2011b} foregrounds how technologies can both reinforce and challenge power relations, and argues for design grounded in pluralism, participation, situated knowledge, and reflexivity~\cite{bardzell2013}. 

Building on these commitments, a substantial body of HCI work has examined how technologies empower (or disempower) women across diverse sociocultural contexts~\cite{kumar2019a, schneider2018, sun2025}, particularly in the Global South~\cite{naseem2020, massimi2012, shroff2011, zhao2024a}. These works foreground the structural nature of oppression, such as online abuse~\cite{sambasivan2019, vashistha2019a, younas2020}, sexual harassment~\cite{ahmed2014a, nova2018}, reproductive injustice~\cite{deva2025a, rahman2021}, and gender-based violence~\cite{rabaan2021, rabaan2023, ndjibu2017, he2025}. For example, Ibtasam~et~al.~\cite{ibtasam2019} show how Pakistani women's access to and use of mobile phones are mediated and restricted by male relatives, while Ahmed~et~al.~\cite{ahmed2024} document how technologies can be misused to violate privacy or constrain women's freedoms in contexts such as forced marriage. This work demonstrates that empowerment through technology is always negotiated within existing power relations rather than simply granted by technological access. In response to these complexities, HCI scholars have proposed design strategies that work with, rather than abstract away from, patriarchal constraints~\cite{sambasivan2019, sultana2018a}. For example, Sultana~et~al.~\cite{sultana2018a} advocate design within a patriarchal society, a pragmatic approach that recognizes that technologies must navigate and sometimes strategically accommodate oppressive structures even while aiming for long-term transformation.

The three bodies of work outlined above provide the foundation for our study. Theoretical scholarship on women's empowerment provides the lens to critically analyze LLM outputs; the Chinese sociopolitical context situates why mild and individualized discourse has become a dominant form of everyday feminism in China; and feminist HCI guides our reflections and design implications.

\subsection{Everyday Digital Feminism}
\label{2.2}
Along with the travel of feminist ideas and awareness, everyday feminism has become a visible branch of feminism that encompasses a wide range of small, individual practices aimed at empowering individuals to challenge gender inequalities in their daily lives~\cite{schuster2017a, kelly2015}. These actions may include calling out a sexist remark among friends, refusing traditional domestic roles, or challenging gendered expectations at work. Social media platforms provide a key environment through which these practices circulate, enabling women to exchange stories, advice, and interpretations at scale~\cite{pruchniewska2019}. In this study, we refer to everyday feminism on social media as \textit{everyday digital feminism}. It represents a strand of digital feminism, but is different from more explicitly political forms of digital activism that directly target institutions or policy~\cite{zeng2019, rodino-colocino2014}.

Building on feminist scholarship, everyday digital feminism can function as an important site of empowerment~\cite{schuster2017a, wang2025b}. First, it creates spaces where women can narrate lived experiences of gendered pressure, violence, or constraint~\cite{batool2022}. Telling one's story publicly or semi-publicly not only makes structural issues visible but also repositions women as speaking subjects, expanding their sense of agency~\cite{wan2025a, stavrositu2008, chang2018}. Second, it provides spaces for women to recognize their struggles in others' accounts, which can counteract shame, isolation, and victim-blaming, thereby cultivating mutual support~\cite{kelly2015, fotopoulou2016, wang2024g, kumari2024}. Moreover, when women are speaking and negotiating, online feminist discourse also helps articulate what counts as ``respect,'' ``abuse,'' ``independence,'' or ``growth'' for women within specific sociocultural contexts~\cite{pruchniewska2019, jouet2018, lazar2019}. In these ways, everyday digital feminist practices contribute to empowerment by shifting who can speak, whose knowledge is treated as legitimate, and which futures feel imaginable. 

In China, RedNote serves as a central site for everyday digital feminism~\cite{wan2025a, guo2022a}, and the widespread discussion of ``women's growth'' is one of the manifestations. Although the term does not carry the explicitly political connotations of empowerment in feminist scholarship, it hosts millions of posts and interactions through which Chinese women articulate concerns, hopes, and negotiations of agency in daily life. Prior research has highlighted the empowering aspects of such everyday feminism on RedNote~\cite{zhan2024, wang2024g, wang2025b}. Yet studies have also shown that the platform is saturated with neoliberal self-optimization discourses, reflected in narratives of entrepreneurialism~\cite{wang2025}, routines for becoming a ``successful woman''~\cite{ge2025}, and self-surveillance practices surrounding the body~\cite{liu2024b, liu2023c}. These discourses both reflect and continually shape how empowerment is framed on RedNote, often emphasizing personal improvement, lifestyle upgrading, and affective fulfillment.

Against this backdrop, the circulation of LLM-generated discourse around ``women's growth'' on RedNote may introduce new interpretations of women's development or reinforce existing ones. It may create both opportunities and risks for everyday feminism as a practice of empowerment. Our study aims to contribute to understandings of these emerging dynamics.

\subsection{Gender Bias in LLMs}
\label{2.3}
Language has long been a primary means through which social norms are produced, but also as a mechanism through which gender inequality is perpetuated~\cite{vanblerck2025, fairclough2013, blodgett2020}. As LLMs have become increasingly prevalent as discursive actors, a growing body of research has examined the gender biases embedded in their outputs~\cite{blodgett2020, gallegos2024, kotek2023}. A substantial literature from the NLP community, using fill-in-the-blank tasks or decision-making scenarios, has revealed that LLMs readily produce gender-related stereotypes concerning occupations, personality traits, and abilities~\cite{dong2023, bartl2025a, gallegos2024}. 

In addition, a growing body of work has examined the normative assumptions about gender embedded in narratives and richer contextual scenarios. For example, Kelly~et~al.~\cite{kelly2025} analyzed 10,000 LLM-generated e-commerce product descriptions on eBay and found gender disparities in body size assumptions, target groups, and persuasive strategies. Some work has offered more critical discourse analyses. Sarikaya~\cite{sarikaya2024} analyzed ChatGPT-generated short stories about family life and found that traditional gendered power dynamics persisted: women were portrayed as responsible for household tasks, while men were more often cast in decision-making or authoritative roles. Sudajit-apa~\cite{sudajit-apa2025} asked ChatGPT to generate 40 ``life-changing'' narrative articles and found that women were framed as powerful, independent, and transformative, yet the narratives rarely provided concrete accounts of how such success could be achieved.

Taken together, prior studies offered understandings of how LLMs portray women and how gendered assumptions surface in generated content. Yet, little is known about how LLMs provide normative advice to women as they navigate their lives, and the underlying assumptions about what women should prioritize, strive for, tolerate, or change. Our work aims to address this gap.
\section{Methods}
\label{methods}
\begin{figure*}[t]
  \centering
  \includegraphics[width=\linewidth]{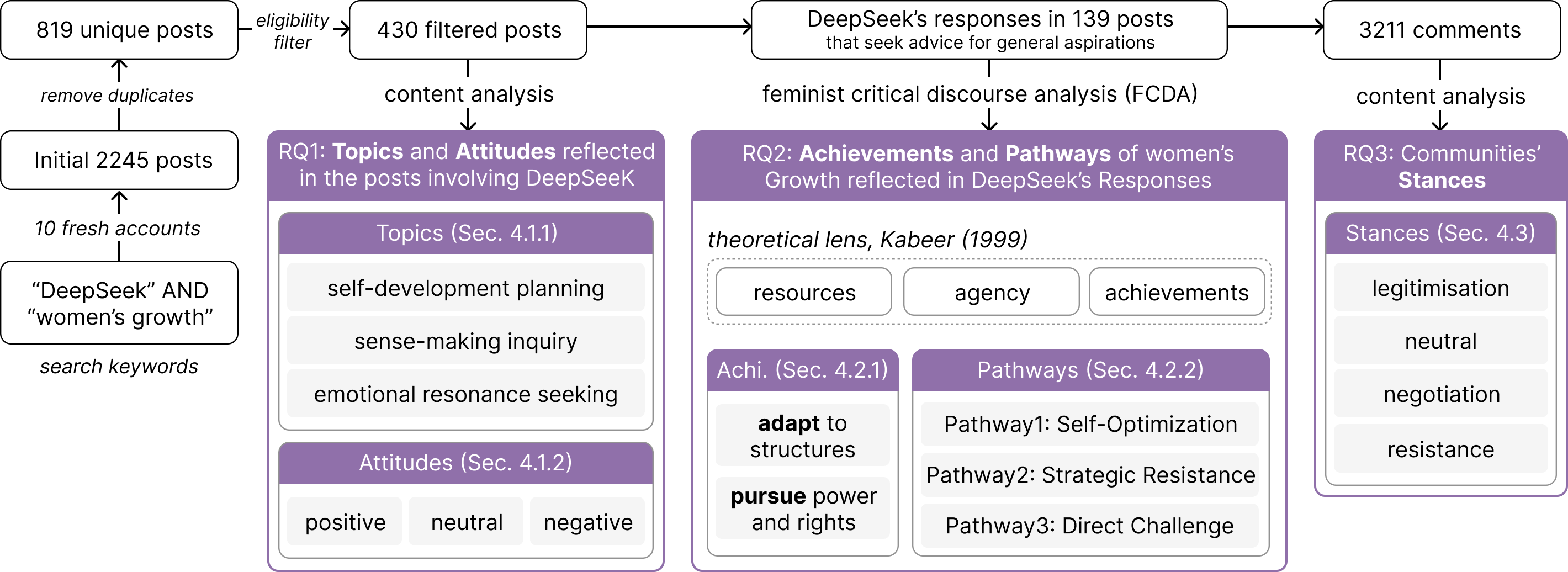}
  \caption{The overall analysis pipeline to investigate how DeepSeek-generated discourse was taken up and evaluated in RedNote}
  \label{fig:methods}
  \Description{Overview of the analysis pipeline used to investigate how DeepSeek generated discourse on ``women's growth'' was taken up and evaluated on RedNote. The figure shows the multi-stage data collection and analysis process. We first collected 2245 posts using the keywords ``DeepSeek'' and ``women's growth,'' then removed duplicates and applied eligibility criteria to obtain 819 unique posts. Content analysis was conducted on 430 filtered posts to address RQ1, which examines the topics and user attitudes in posts involving DeepSeek. Three topics were identified: self-development planning, sense-making inquiry, and emotional resonance seeking, alongside positive, neutral, and negative attitudes. For RQ2, feminist critical discourse analysis was applied to DeepSeek's responses in 139 posts that sought general aspiration advice. Using Kabeer's framework of resources, agency, and achievements, we identified two types of achievements (adapting to structures and pursuing power and rights) and three pathways of women's growth reflected in the responses: self-optimisation, strategic resistance, and direct challenge. For RQ3, content analysis of 3211 comments identified four community stances toward DeepSeek's advice: legitimisation, neutral, negotiation, and resistance. The figure illustrates how each analytical step builds on the previous one to show how DeepSeek's discourse was produced, circulated, and evaluated within the feminist digital space.}
\end{figure*}

To systematically investigate how DeepSeek-generated content is being incorporated into public discussions of women's growth on RedNote, we collected relevant posts, DeepSeek responses included, and comments as the corpus, and conducted a series of qualitative analyses. The overall analysis pipeline is illustrated in \autoref{fig:methods}.

\subsection{RedNote as the Data Source Platform}
We selected RedNote as our data source, which has also been examined in prior research within and beyond HCI on feminist topics~\cite{wan2025a, wang2025, liu2023c, pan2025a}. On the one hand, it hosts a large and active user base. According to \cite{hau2024}, RedNote has around 100 million monthly active users, most of whom are between 18 and 35 years old, with women comprising about 70\% of the audience. This widespread popularity and demographic composition make it a highly influential arena for public discussion. On the other hand, as noted in \autoref{2.2}, RedNote has become a central site for digital feminism in China, making it a particularly promising context for examining how emerging technologies intersect with feminist conversations.  

\subsection{Data Collection and Processing}
Since DeepSeek's chatbot release in January 2025, discussions about its use have increasingly appeared on RedNote, including frequent engagements with women's growth-related topics. We first noticed this phenomenon in January 2025 and conducted continuous observations through May 2025, during which we observed a steady increase in both the volume and diversity of relevant content. We conducted systematic data collection in May 2025 using an open-source tool, MediaCrawler\footnote{See \url{https://github.com/NanmiCoder/MediaCrawler}}.

Before large-scale collection, we conducted empirical observations and preliminary crawling to examine how RedNote's search and content retrieval mechanisms operate. Using the keywords ``women's growth'' AND ``DeepSeek,'' we observed that the platform does not return results in a strictly ranked order (\eg by likes). Instead, it appears to apply a weighting mechanism that surfaces posts with varying levels of engagement, resulting in different sets of posts across repeated queries. This ensured that our dataset captured content across a range of popularity levels. Consistent with observations in \cite{deng2025}, our testing indicated that RedNote's algorithm returns not only posts containing the exact queried keywords but also content with semantically related themes. For example, searches for ``women's growth'' (女性成长) also surfaced posts mentioning ``women's power'' (女性力量), ``women's awakening'' (女性觉醒), and ``female independence'' (女性独立). We retained such posts during initial data curation to enrich the diversity and representativeness of our dataset. 

Based on observations from the preliminary crawling, we proceeded to the main data collection. We created ten fresh RedNote accounts without prior activity for data collection, as our pilot testing showed that RedNote returns partially different results across accounts. Using multiple accounts provided more coverage of relevant posts, a practice also adopted in prior HCI research~\cite{deng2025}. We used the platform's default sorting and collected all available content types associated with each post, including text, images, and videos. In total, we gathered 2,245 posts; after removing duplicates, we obtained 819 unique entries. \camera{It is worth noting that, although the scraping was conducted within a data collection window in May 2025, MediaCrawler enabled us to retrieve posts published prior to the search. As a result, our corpus covers content published between January and May 2025, spanning from DeepSeek's launch to the point at which systematic data collection was carried out.} 

To remove irrelevant posts, two authors independently screened 100 randomly sampled posts to develop inclusion and exclusion criteria. After discussion, we finalized the criteria and applied them to the full dataset. We excluded 389 posts for the following reasons: (1) not related to women's growth (n=170); (2) no interaction with DeepSeek (n=57); (3) advertisements (n=28); (4) prompt-sharing posts (n=78, \eg ``\textit{must-know prompts for using DeepSeek}''); and (5) general self-development inquiries not specific to women (n=56, \eg ``\textit{asking DeepSeek how to learn swimming}''). This yielded 430 posts in which users engaged with DeepSeek in discourses relevant to women's growth. Among the 430 posts, 247 were by female users (57.4\%), 26 by male users (6.0\%), and 163 by users who did not specify gender (36.6\%). 

\subsection{Data Analysis}
\subsubsection{Topics and Attitudes in Posts (RQ1)}
\label{rq1_methods}
We conducted a content analysis of 430 post titles, descriptions, and screenshots of prompts to DeepSeek, where available, to examine the topics related to women's growth and the attitudes expressed toward DeepSeek responses. Three authors independently open-coded an initial random sample of 50 posts, generating preliminary codes along two dimensions: (1) types of conversations and topics related to women's growth (\eg \textit{general aspirations}, \textit{intimacy relations}); and (2) posters' attitudes when sharing DeepSeek responses (\eg \textit{positive}, \textit{negative}). The coders then reconciled differences to develop a shared codebook. This codebook was applied to additional batches of 50 posts and refined iteratively, with new codes added or merged as needed. Thematic saturation was reached after coding approximately 200 posts. Each post was then independently coded by two coders, with disagreements resolved in weekly meetings. The findings are presented in \autoref{fig:fullsankey}. 

\subsubsection{Feminist Critical Discourse Analysis (FCDA) of DeepSeek-generated Content (RQ2)}
\label{rq2_methods}
\textbf{Corpus}: To address RQ2, we first narrowed down our corpus by focusing on the 313 posts categorized as \textit{Self-Development Planning} (see \autoref{fig:fullsankey}). Compared with other categories, this subset was distinctive in that the responses provided actionable guidance, combining suggested achievements with concrete pathways for women's growth. This made it the most suitable corpus for examining how DeepSeek framed what women should aspire to and how they might pursue those goals. Within this category, we then concentrated on a further subset of 139 posts labeled as \textit{General Aspirations}. Since these questions were especially open-ended, asking broadly about how women can live well or find meaning in their lives without specifying particular domains or outcomes, they offered greater space to observe how DeepSeek articulated women's achievements and pathways. The final corpus for RQ2, therefore, consisted of DeepSeek responses from 139 posts. After transcription, these responses amounted to a total of 114,162 Chinese characters, which served as the basis for RQ2 analysis. 

\textbf{Methodology and Theoretical Lens}: We adopted feminist critical discourse analysis (FCDA)~\cite{lazar2014} as our methodological approach to analyze these responses. Building on critical discourse analysis~\cite{fairclough2023}, FCDA examines how gender, power, and ideology intersect in language and communication to create and perpetuate gendered social structures. In our study, where LLMs such as DeepSeek produce and mediate discourses about gender, FCDA is particularly relevant. It provides a critical lens for analyzing how these discourses reflect gendered power dynamics, as also demonstrated in prior HCI research using this approach~\cite{qin2024,lazar2019}. 

To be more specific, we used Kabeer's framework on women's empowerment~\cite{kabeer1999} as our analytic lens. As detailed in \autoref{2.1}, the framework conceptualizes resources as preconditions, agency as the capacity to act, and achievements as outcomes. Accordingly, we operationalized these three components as identifiable discursive elements in DeepSeek's advice. We first coded \textit{achievements}, which refer to the outcomes depicted as desirable in the responses, including the life domains they relate to and whether they adapt to or challenge existing structures. For each achievement, we then examined its associated \textit{resources}, beginning with whether any enabling conditions were mentioned and, when present, coding their types (such as skills, financial means, networks, or legal channels). We coded \textit{agency} when the text prescribed forms of action, including self-regulation, planning, negotiation, boundary setting, or collective action. Finally, by analyzing how resources, agency, and achievements were linked within each piece of advice, we identified the pathways suggested in DeepSeek's responses. 

\textbf{Detailed Procedures}: All texts were analyzed in Chinese without translation to preserve linguistic features and context. Two authors served as coders and began with close readings of the corpus to familiarize themselves with the contexts, vocabularies, and narrative structures in DeepSeek responses. Next, responses from 20 posts were randomly selected for the coders to independently identify themes and patterns. Through discussion, they interpreted the quotations with attention to how they reinforced or challenged gender norms and power relations, developed explanations of these dynamics, and constructed an initial codebook covering achievements, agency, and resources. The refinement of the codebook followed the same iterative procedure as in the RQ1 analysis (see \autoref{rq1_methods}): batches of responses from 20 posts were coded, disagreements were resolved in weekly meetings, and other co-authors participated to help mediate potential bias. In total, the analysis yielded 842 coded achievements, which were grouped into two categories and multiple domains (see \autoref{fig:RQ2.1}). By examining the linkages among resources, agency, and achievements, we further identified three recurring pathways (see \autoref{fig:RQ2.2}). 

Lastly, we acknowledge that DeepSeek's outputs can vary depending on the specific prompts or preceding interaction turns. Given our research focus on how its responses circulated and were interpreted within RedNote, we analyzed the user-shared excerpts as they appeared on the platform. This approach is appropriate for our research questions, which concern the discourse that RedNote users actually encountered and engaged with. Thus, we treated each shared response as an independent discursive instance while noting that our findings do not represent DeepSeek's full behavior under controlled settings.

\subsubsection{Content Analysis to Identify Commenters' Stances (RQ3)}  
\label{rq3_methods}  
To address RQ3, we focused on first-layer comments as they directly capture audience reactions to the post content (\ie DeepSeek's suggestions). We retrieved the comments under the 139 posts analyzed in RQ2, which yielded 3,417 entries. After removing irrelevant comments such as advertisements and off-topic content (n=206), we analyzed 3,211 eligible comments. Among them, 2,325 were by female users (72.4\%), 128 by male users (4.0\%), and 758 by users who did not specify gender (23.6\%). 

The coding aimed to identify the stances reflected in the comments (\eg \textit{legitimization}, \textit{negotiation}, \textit{resistance}, or \textit{neutral}) and the specific forms through which these positions were expressed (\eg \textit{emotional resonance}, \textit{praise for practicality or applicability}). The analysis considered both the context of the original posts and the corresponding DeepSeek responses. All comments were independently coded by two coders, following the similar procedures used in RQ1 (see \autoref{rq1_methods}), with regular meetings held to resolve disagreements. The findings are presented in \autoref{fig:RQ3}.  

\subsection{Positionality Statement}  
\subsubsection{Positionality Statement}
Following the principles of FCDA and the reflexive tradition of feminist HCI studies~\cite{bardzell2010a, bardzell2011b}, we reflected on our positionalities that shaped how we approached, interpreted, and analyzed the data. 

All authors are Chinese researchers engaged in feminist HCI scholarship and practice. We recognize the long history of gender inequality and advocate for women's empowerment and gender equality through individual, collective, and structural efforts with and against technologies. This commitment shaped the foundational motivation of the present work. The data analysis and interpretation process included five cisgender female researchers in their 20s to 40s and two cisgender male researchers in their 20s and 50s, all raised in China. These backgrounds brought diverse perspectives for our understanding of the sociocultural meanings associated with ``women's growth'' in the Chinese context. In addition, we consulted with feminists outside academia to further reflect on our interpretations of DeepSeek's responses. At the same time, we acknowledge the limitations of our standpoint. Because all authors have received higher education and have lived primarily in major cities in China, our interpretations may not fully represent the perspectives of women from rural areas or with limited literacy.

\subsubsection{Ethics Considerations}  
All data analyzed in this study were collected from publicly accessible posts and comments on RedNote. Before analysis, the materials were anonymized and desensitized to remove any user identifiers. We present only de-identified excerpts and translated quotes. The study protocol was reviewed and approved by our institutional ethics committee (IRB).

\section{Findings}
In this section, we first present the \textbf{\textit{topics}} on women's growth raised to DeepSeek and the \textbf{\textit{attitudes}} reflected in their posts (RQ1, \autoref{rq1}). Second, we show the \textbf{\textit{achievements}} envisioned and the \textbf{\textit{pathways}} DeepSeek suggested to women (RQ2, \autoref{rq2}). Finally, we present the different \textbf{\textit{stances}} expressed toward DeepSeek suggestions (RQ3, \autoref{rq3}). Below, we present translated examples from the corpus. Please refer to the supplementary materials for the original Chinese text, the English translations, and the full codebooks for each RQ.

\subsection{RQ1: Topics and Attitudes in Posts}
\label{rq1}

\begin{figure*}[t]
  \centering
  \includegraphics[width=\linewidth]{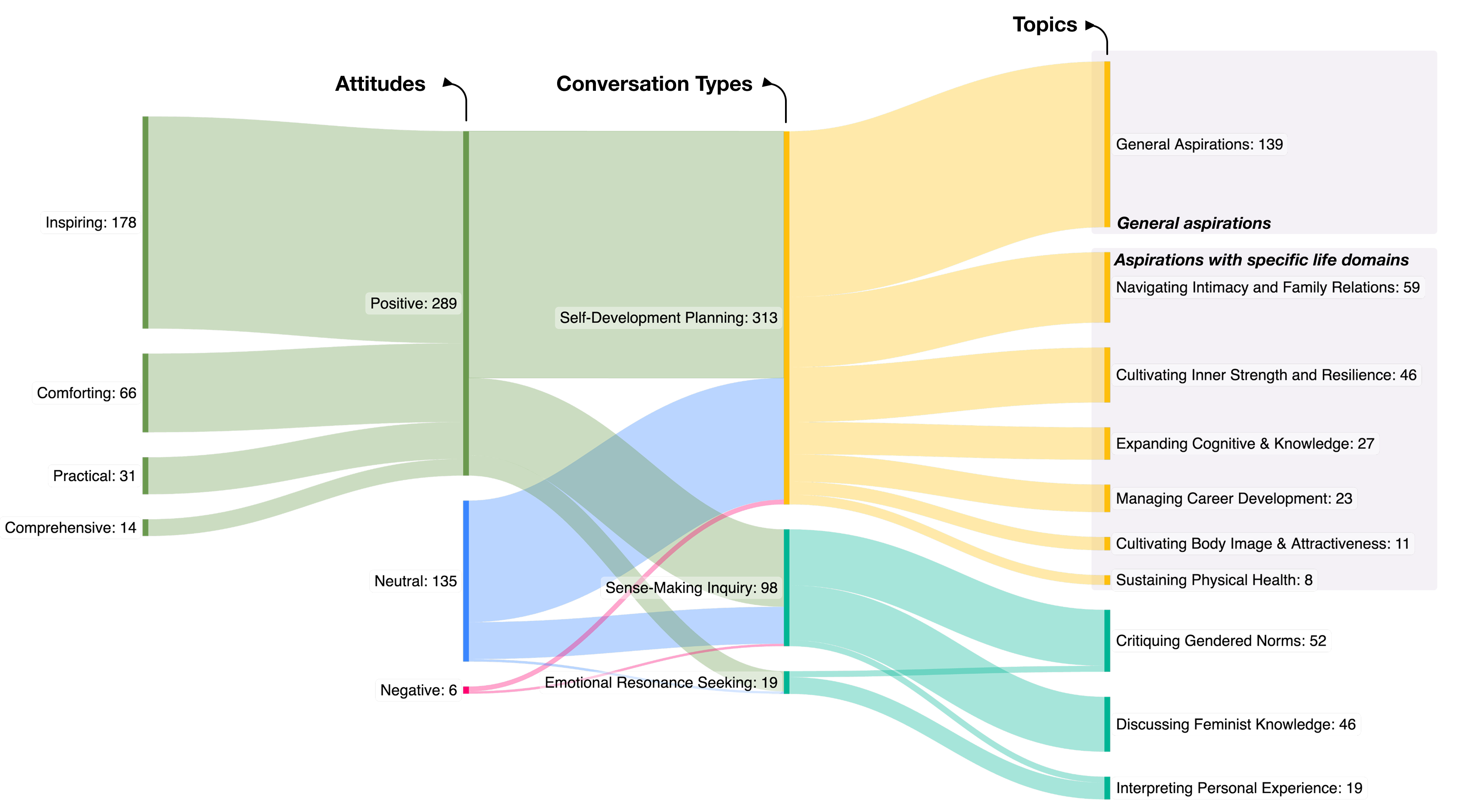}
  \caption{The visualization of RQ1 findings. It shows how RedNote posts framed DeepSeek in relation to women's growth, moving from expressed attitudes, to the conversation types they engaged in, and finally to the topics identified in their conversations.}
  \label{fig:fullsankey}
  \Description{A Sankey diagram visualizing the codebook of RQ1, showing how RedNote creators framed DeepSeek in relation to women's growth. On the left, attitudes are categorized as inspiring (178), comforting (66), practical (31), comprehensive (14), neutral (135), and negative (6). These flow into conversation types in the middle, including self-development planning (313), sense-making inquiry (98), and emotional resonance seeking (19). On the right, the flows end in topics such as general aspirations (139), navigating intimacy and family relations (59), cultivating inner strength and resilience (46), expanding cognitive knowledge (27), managing career development (23), cultivating body image and attractiveness (11), sustaining physical health (8), critiquing gendered norms (52), discussing feminist knowledge (46), and interpreting personal experience (19). The diagram illustrates how initial attitudes shape types of conversations with DeepSeek and ultimately connect to broader aspirations and discursive orientations around women's growth.}
\end{figure*}

\subsubsection{Topics Raised to DeepSeek} 
We identified that users raised various \textit{topics} regarding women's everyday life (see \autoref{fig:fullsankey}). We categorized them into three types of conversations: 

\textbf{\textit{Seeking Self-Development Suggestions}} (313 instances, 72.8\%). This was the predominant conversation type, where DeepSeek was positioned like an \textit{action planner} that offered suggestions for navigating diverse aspirations. Across these posts, we observed that some questions focused on \textbf{Aspirations within Specific Life Domains}, such as marriage, career, or health. As shown in \autoref{fig:fullsankey}, a substantial subset focused on intimacy and family relations, seeking advice on decisions about marriage, reproduction, or partnership, such as ``\textit{What is the optimal marriage age for women according to DeepSeek?}'' Others foregrounded inner psychological development, asking how to build emotional stability, or emphasized cognitive and knowledge growth, including feminist learning (e.g., \textit{``10 tips for women beginning their feminist journey''}). Less frequent were posts that situated women's growth in career advancement, body image and attractiveness, or sustaining physical health across different life stages. More examples can be found in \autoref{appendix-rq-1.1}.

In contrast, other posts raised broad, open-ended concerns about women's growth without anchoring them to any particular life domain. We refer to these domain-agnostic questions as \textbf{General Aspirations}. As shown in \autoref{fig:fullsankey}, it emerged as the most common category. In these posts, users posed open-ended questions about women's life trajectories and self-development. For example, \textit{``How can women live a life without regrets?''} or \textit{``How can ordinary girls quickly improve themselves?''} Such inquiries provided DeepSeek with interpretive space not only to suggest action plans but also to define what counts as valuable goals to pursue. 

\textbf{\textit{Sense-Making Inquiry}} (98 instances, 22.8\%). 
The second major conversation type involved content creators showcasing DeepSeek's responses around critiquing gendered norms (e.g., \textit{``What are the top ten distorted beauty standards that constrain women?''}), discussing feminist knowledge (e.g., \textit{``Please summarize Chizuko Ueno's feminist ideas''}), and interpreting personal experience (e.g., \textit{``Why do I, as a self-identified independent woman, enjoy reading Stepford Wives-style stories?''}). In this category, DeepSeek functioned as a \textit{discussion partner} that provided interpretive insights and discussions. 

\textbf{\textit{Seeking Emotional Resonance}} (19 instances, 4.4\%). 
A smaller group of posts sought emotional validation or empathy from interactions with DeepSeek. They recounted personal struggles and used DeepSeek's responses as a source of comfort or encouragement. For example, one video post featured a girl who shared her feelings of low self-esteem due to body stigma. As she read DeepSeek's reply aloud, she was moved to tears and described the response as deeply comforting. Although less frequent, these conversations positioned DeepSeek as an \textit{empathetic companion}.

\subsubsection{Attitudes Toward DeepSeek Responses}
As shown in \autoref{fig:fullsankey}, we categorize the attitudes reflected in the posts as positive, negative, or neutral, with positive framing being predominant. 

\textbf{\textit{Positive Attitudes}} (289 instances, 67.2\%).  
The majority of posts expressed positive attitudes toward DeepSeek's responses, positioning DeepSeek as a valuable contributor to discussions of women's growth. Most of these posts framed DeepSeek as \textit{inspiring}: users described its answers as ``\textit{eye-opening}'' or giving them ``\textit{a completely new perspective}.'' Some encouraged others that \textit{``in times of internal conflict, talk to DeepSeek; whether it is about a life plan or a woman's dilemma, it will analyze everything for you.''} Positive posts also frequently emphasized the \textit{comforting} role of DeepSeek, highlighting its emotional support. One user wrote, \textit{``DeepSeek is actually an emotional comfort device. I'm a very sensitive person who rarely talks with people around me, but when I was completely overwhelmed with nowhere to vent, I told it everything.''} A smaller portion of posts praised the \textit{practical} and \textit{comprehensive} feature of DeepSeek's suggestions, describing them as detailed, executable plans that could be implemented step by step.

\textbf{\textit{Negative Attitudes}} (6 instances, 1.4\%).  
In contrast, only a few posts voiced explicitly negative views. Four criticized DeepSeek for reproducing gender bias in career advice, noting that it assumed women should balance family and work. For example, one user remarked, \textit{``I don't really agree with DeepSeek's answer this time. Its answer is consistent with workplace discrimination against women. It naturally assumes women should balance family and work, and be delicate and gentle in all life scenarios.''} Two others critiqued its relationship advice, observing that when a woman described being in a toxic relationship, DeepSeek suggested she \textit{``give him space,''} whereas men in similar scenarios were advised by DeepSeek to \textit{``document evidence''} and \textit{``contact authorities''}. 

\textbf{\textit{Neutral Attitudes}} (135 instances, 31.4\%).  
Still, a substantial portion of posts did not reflect explicit attitudes, often accompanied by little or no additional commentary from the creators.  

% Together, these findings address RQ1, showing that DeepSeek was positioned as an action planner, discussion partner, or empathetic companion in relation to various aspects of women's growth. While the predominant attitude toward DeepSeek was positive, some posts explicitly highlighted its gender bias. 

% examples: 

% 当我问deepseek：30+的女性如果想在一二线城市获得一份职业成就，可以在哪些行业取得比较好的发展？其实这次deepseek的答案我并不十分赞同，它的思考过程和现在的职场社会如出一致，充斥着对女性的职场歧视——它天然地认为女性应该平衡家庭和职场，女性应该是细腻温柔的……

% comforting: DeepSeek其实是一款情绪安慰器. 我是很敏感的人，平时不会和身边人说太多话，但今天崩溃得实在没处宣泄的时候和它说了很多，后来突然想知道ai会不会对我感到厌烦，会不会和人一样不愿意接#受别人的负面情绪; DeepSeek给35+女性的建议，看哭了～ 
% practical: 关于已婚已育女性成长DeepSeek给的很实用; 当我问deepseek如何用一年时间从普女逆袭成气质美女，它给出的方案可执行性也太太太强了吧！
% comprehensive: 还是太全面了点！; 今天我通过与DeepSeek的对话，给我一个输出了非常全面的言说女性成长规划与发展的计划
% inspiring: 都说《被讨厌的勇气》是神书 看懂的缺不到1% 我问deepseek 到底讲了啥？ 对我们有什么启示？ deepseek的拆解， 我感觉要开窍了; 姐妹们！在内耗的时候，可以找DeepSeek聊聊啊 不管是人生规划还是女性困境，保准给你分析地明明白白

% 如何建立女性主义知识体系？
% 女性人生投产比最高的18件事？
% 如何做个情绪稳定的妈妈？
% 哪些先进的行业中女性比男性更具有优势？如何进入这些行业？

\subsection{RQ2: Achievements Envisioned and the Pathways DeepSeek Suggested to Women}
\label{rq2}
As detailed in \autoref{rq2_methods}, we used Kabeer's framework~\cite{kabeer1999} to conduct the feminist critical discourse analysis. In this section, we first present the \textit{achievements} identified, then move to the \textit{pathways} that DeepSeek appeared to set for women to pursue these achievements.   

\subsubsection{Achievements Envisioned by DeepSeek}
\label{rq2.1}
\begin{figure*}[t]
  \centering
  \includegraphics[width=\linewidth]{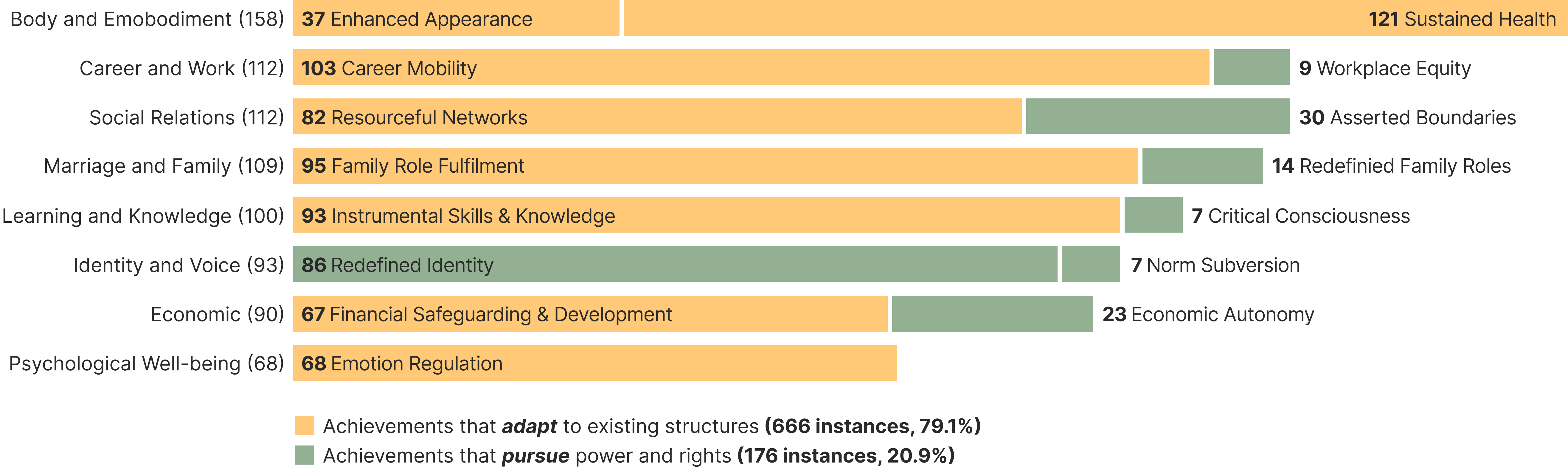}
    \caption{Categories of achievements reflected in DeepSeek's responses, as shared on RedNote. Achievements that \textbf{\textit{adapt}} to existing structures accounted for the majority (666 instances, 79.1\%). These emphasize performance, accumulation, and competitiveness without questioning systemic inequalities. Achievements that \textbf{\textit{pursue}} power and rights (176 instances, 20.9\%) focus on resisting gender norms, redefining family roles, and pursuing equity in public and professional life.}
    \Description{A bar chart showing categories of achievements reflected in DeepSeek's responses on RedNote, divided into two groups: achievements that adapt to existing structures (orange, 666 instances, 79.1\%) and achievements that pursue power and rights (green, 176 instances, 20.9\%). Categories include body and embodiment (158, with 37 enhanced appearance and 121 sustained health), career and work (112, with 103 career mobility and 9 workplace equity), social relations (112, with 82 resourceful networks and 30 asserted boundaries), marriage and family (109, with 95 family role fulfillment and 14 redefined family roles), learning and knowledge (100, with 93 instrumental skills and knowledge, 7 critical consciousness), identity and voice (93, with 86 redefined identity, 7 norm subversion), economic (90, with 67 financial safeguarding and development, 23 economic autonomy), and psychological well-being (68, with 68 emotion regulation). The visualization shows that most responses emphasized adaptation through performance, accumulation, and competitiveness, while a smaller proportion focused on resisting gender norms, redefining family roles, and pursuing equity in public and professional life.}
  \label{fig:RQ2.1}
\end{figure*}

As shown in \autoref{fig:RQ2.1}, we identified achievements in different life domains, such as career and work, social relations, and marriage and family. Moreover, we further identified two distinct discursive orientations: whether they \textbf{\textit{adapt to}} or \textbf{\textit{challenge}} existing structures of gender inequality. Most of the identified achievements fall into the former, adapting to existing structures. We detail the findings below, and use \textit{italics} to indicate quotations translated from DeepSeek. 

\textbf{\textit{Achievements that Adapt to Existing Structures}} (666 instances, 79.1\%) dominated nearly all domains except the Identity and Voice (see \autoref{fig:RQ2.1}). These achievements envisioned women's development across various life domains without explicitly acknowledging structural barriers underpinning gender inequality.

For example, in the \textbf{Career and Work}, the vast majority of suggestions (103 out of 112) emphasized upward mobility, such as ``\textit{being highly competitive},'' but they seldom mentioned goals such as leadership or political authority that have traditionally been male-dominated. In the \textbf{Economic} domain, most advice (67 out of 90) encouraged women to ``\textit{be economically capable},'' pursue ``\textit{systemic income},'' or ``\textit{run side businesses}.'' Yet, the proposed ventures often reflected gendered expectations, such as ``\textit{monetizing hobbies in drawing or crafting}'' or ``\textit{working as social media creators in fashion}.'' In the \textbf{Marriage and Family}, where gendered expectations are deeply entrenched~\cite{zhao2024a, yang2025}, DeepSeek often (95 out of 109) assumed women would enter marriage by default. The envisioned achievements, hence, centered on reducing marital risks, such as ``\textit{securing a prenuptial agreement},'' and fulfilling service roles within the family, such as ``\textit{reliable carriers sustaining family operations}''. 

Besides, a strong instrumental orientation was observed in this category. In the \textbf{Social Relations}, most goals (82 out of 112) stressed cultivating ``\textit{useful connections}'' and ``\textit{high-quality ties},'' emphasizing utility rather than solidarity or care. In the \textbf{Learning and Knowledge}, women were advised to become ``\textit{erudite individuals},'' framed as ``\textit{horizon-broadened, highly adaptable professionals}'' or ``\textit{capable communicators},'' with little reference to learning for intrinsic interest or enjoyment of knowing. 

\textbf{\textit{Achievements that Pursue Power and Rights}} (176 instances, 20.9\%) explicitly questioned, resisted, or sought to transform the norms sustaining gender inequality. Although comparatively few, such achievements surfaced across multiple domains. Notably, the domain of \textbf{Identity and Voice} featured suggestions for reconstructing women's roles and self-conceptions. For example, DeepSeek cautioned, \textit{``Be wary of `social scripting.' Traditional definitions of women's worth (marriage, childbearing, devotion) are role expectations, not life answers. True `living life to the fullest' begins with stripping away the external noise and listening to the heart: `What do I want? What makes me feel truly alive?'}'' Beyond defining oneself, a few suggestions encouraged women to act publicly as \textit{``feminist advocates''} or \textit{``social activists who fight for women's rights.''} In the domain of \textbf{Career and Work}, a handful of responses explicitly encouraged women to ``\textit{recognize workplace gender discrimination}'' and to reposition themselves from ``\textit{rule adapters to rule makers}.'' In the \textbf{Economic} domain, the envisioned achievement of ``\textit{economic autonomy as the key to life choices}'' stood out as a departure from the more common framing of income stability. Finally, in the \textbf{Marriage and Family}, a small set of suggestions urged women to resist entrenched labels such as ``\textit{good wife and loving mother}'' or ``\textit{obedient daughter},'' pointing to the possibility of redefining familial roles. 

%警惕``社会脚本''的绑架 传统对女性价值的定义（婚姻、生育、奉献）本质是角色期待，而非人生答案。真正 的``不枉此生''应始于剥离外界噪音，倾听内心：``我想要什么？什么让我真正感到活着？''

%经济独立才能拥有更多人生选择权 
% 撕掉``贤妻良母''``乖巧女儿''的标签

% 拒绝被年龄定义、``拒绝社会标签''、``打破社会规训的束缚''. "从``适应规则''到``制定规则''：女性不必仅满足于在现有体系中生存，可以成为领导者、创业者，重新定义游戏规则。打破``雌竞''叙事，共同争取更大话语权``直面野心：在职场中敢于争取机会，表达诉求'' "接纳情绪流动：悲伤、愤怒、失落都是正常的，不必强迫自己``积极''。" 

%``身体是革命的本钱''. 
%有风险意识的
% ``建立「财务防火墙」，强制储蓄工资的20%...开设独立应急账户，存够3-6个月生活费，避免因突发状况陷入被动''
% 强调``婚前协议很必要''、``婚前查对方的征信记录''等；另一方面，DeepSeek 建议进入婚姻的女人要认真经营婚姻，例如``婚姻需要``合伙人思维''''、``家庭运营：建立「航母式支持系统」''
%``打造可迁移的核心竞争力''、``完成职位上需要工作和让女人提高竞争力、and seeking new opportunities 。但通常很少提到要让女人站上管理层、成为专家等。相似地，在Economic方面，DeepSeek 虽然建议女性展开副业来创造元多收入，但是大多数提到的技能都偏向于传统的女性角色的技能，比如``将爱好（如插画、手作）转化为收入''、或者是``运营小红书/抖音（分享擅长领域：穿搭/读书/职场）''。很少提到XXX。
% 另外的观察是，大多数这个类别下的成长建议，都带有比较明确的instrumental的倾向。比如，不论是学习还是社交、还是处理情绪。例如，在Learning and Knowledge方面，女人被envision成要博学广识：``构建认知护城河''、``认知升级''、学习的目的多被强调为``拓宽视野，提升应变力''、``实现思维输出''、``''培养表达力``等，而非兴趣、爱好。相似地，在人际交往方面，DeepSeek 强调女人的achievements是``战略性地建立人脉池''、``建设优质社交网络''。强调社交的目的性。

\subsubsection{Pathways Suggested by DeepSeek to Pursue the Achievements} 
\label{rq2.2}
\begin{figure*}[t]
  \centering
  \includegraphics[width=\linewidth]{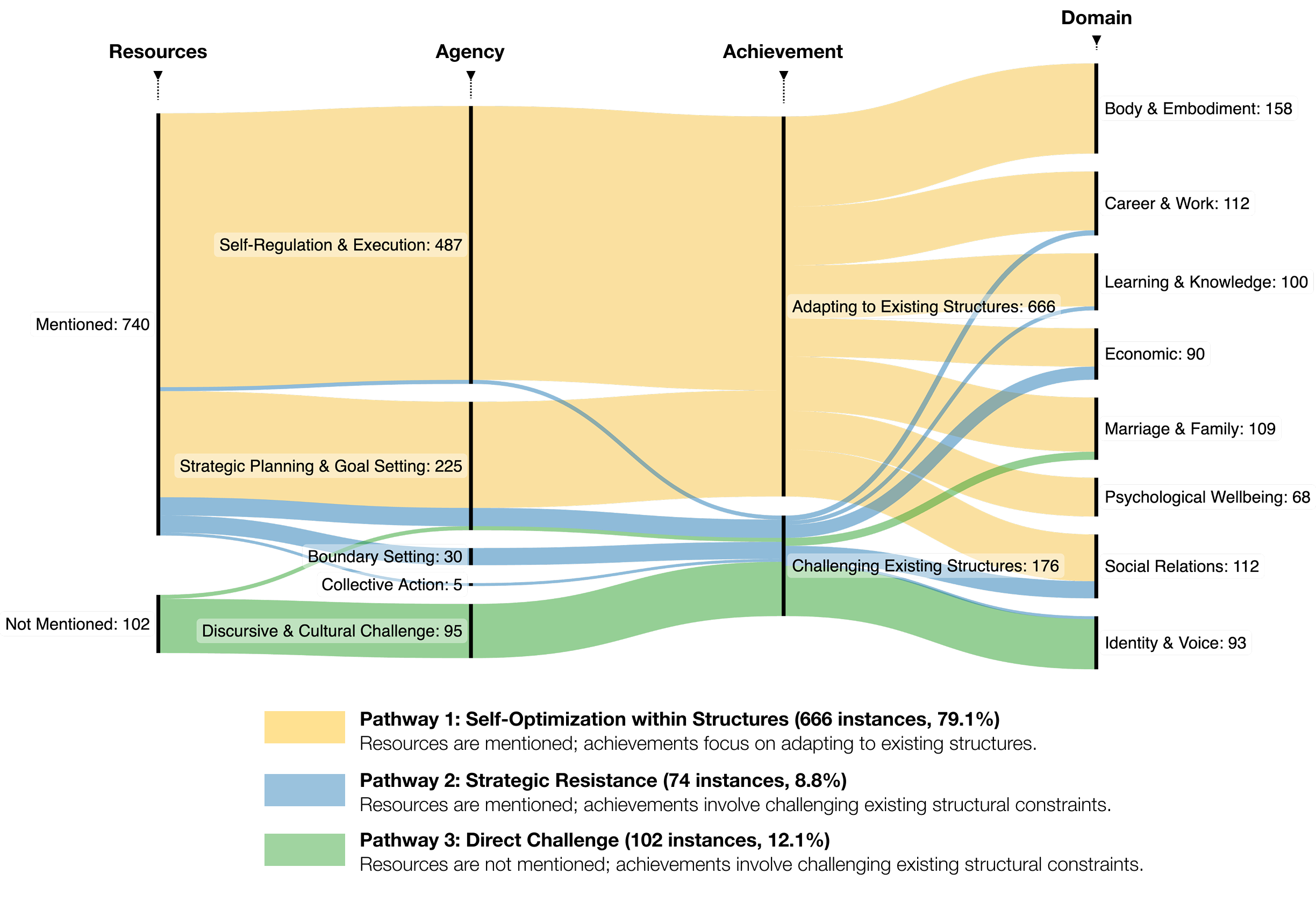}
    \caption{Three pathway patterns identified in DeepSeek responses. The Sankey diagram shows the flow from \textbf{resources} (left) through \textbf{agency} (center) to \textbf{achievement} and life domains (right). Based on the flow patterns, three distinct pathways of envisioned achievements emerged in DeepSeek responses: Self-Optimization within Structures (Pathway 1), Strategic Resistance (Pathway 2), and Direct Challenge (Pathway 3).}
    \Description{A Sankey diagram illustrating three pathway patterns in DeepSeek responses, tracing the flow from resources (left) through agency (center) to achievement and life domains (right). Of 842 instances, 740 mentioned resources, most prominently self-regulation and execution (487), strategic planning and goal setting (225), boundary setting (30), collective action (5), and discursive and cultural challenge (95). These are connected to two main achievement types: adapting to existing structures (666 instances, 79.1\%) and challenging existing structures (176 instances, 20.9\%). Life domains on the right include body and embodiment (158), career and work (112), learning and knowledge (100), economic (90), marriage and family (109), psychological well-being (68), social relations (112), and identity and voice (93). The pathways are color-coded: Pathway 1 (yellow, 666 instances, 79.1\%) represents self-optimization within structures, Pathway 2 (blue, 74 instances, 8.8\%) represents strategic resistance, and Pathway 3 (green, 102 instances, 12.1\%) represents direct challenge. The figure highlights how most responses emphasized individual adaptation and optimization, while smaller portions engaged in resistance or direct challenges to systemic structures.}
  \label{fig:RQ2.2}
\end{figure*}

For each identified achievement, we further examined how resources and agency were articulated, using Kabeer's framework~\cite{kabeer1999} as an analytic lens. The framework emphasizes that empowerment emerges from the interplay among resources, agency, and achievements. Accordingly, our analysis focused on the relationships among these three components. We derived three recurring patterns, which we refer to as pathways (see \autoref{fig:RQ2.2}). Below, we describe each pathway, attending in each case to (1) the resources that were or were not made available, (2) the forms of agency encouraged, and (3) the kinds of achievements envisioned.

% Drawing on Kabeer's framework~\cite{kabeer1999} of resources, agency, and achievements, we examined how these three dimensions were reflected in DeepSeek's suggestions. 

%According to each achievement identified, we first identified the types of actions prescribed and interpreted these as forms of \textit{agency}. Through this step, we identified five distinct forms of agency, such as self-regulation and collective action. We then examined whether \textit{resources} were explicitly mentioned as preconditions for these actions or for the envisioned achievements. 

% Resources → Adapting: 666 (79.1%)
% Resources → Challenging:  74 (8.8%)
% Not mention → Adapting:   0 (0.0%)
% Not mention → Challenging: 102 (12.1%) 

\textbf{\textit{Pathway 1: Self-Optimization within Structures}} (666 instances, 79.1\%) dominated the corpus (see \autoref{fig:RQ2.2}), spanning nearly all life domains except the \textbf{Identity and Voice}. DeepSeek responses following this pattern typically presented \textit{resources} (\eg skills, certifications, or social capital) as attainable targets through individual effort. \textit{Agency} types were overwhelmingly concentrated in self-regulation and execution (480 instances out of 666), as well as strategic planning and goal setting (186 out of 666). The envisioned \textit{achievements} focused exclusively on adapting to existing structures. 

These discourses assumed that resource investment and disciplined self-management through strategic planning and goal setting (186 out of 666, see \autoref{fig:RQ2.2}) would directly lead to success, while obscuring systemic barriers that may impede women's advancement. For example, one response envisioned career mobility as the promised achievement. The corresponding advice explicitly listed resources such as technical skills and professional certifications as acquisitions obtainable through individual effort and planning: 

\begin{quote}  
    \textit{``Construct a personal competitiveness matrix by using Excel to document hard skills such as Python or financial modeling, soft skills such as project management or cross-departmental communication, and industry certifications such as CFA or PMP, then develop an annual plan for improvement, ..., to be promoted to project manager within three years and increase salary by 70\%.''}
\end{quote}  

By framing career success as a straightforward equation of skill accumulation and personal discipline, this suggestion omits structural barriers such as gender discrimination in promotion decisions. Similar cases can be observed in the \textbf{Marriage and Family} domain, where DeepSeek's responses presupposed that women will get pregnant and suggested that through strategic planning and reproductive technologies, women can better manage reproduction:

\begin{quote}
    \textit{``If you don't have a reliable marriage partner by age 35, immediately activate egg freezing or single motherhood plans. Don't wait until menopause to regret it.''}
\end{quote}  

However, the potential barriers, such as the high economic cost of fertility treatments, are not mentioned here~\cite{li2025b}. Besides, it inadvertently associates reproduction solely with women and ties it closely to age, which also promotes reproductive planning as an individual responsibility and perpetuates anxieties about aging. 

Additionally, a large number of suggestions emphasized self-regulation and execution (480 out of 666, see \autoref{fig:RQ2.2}) as pathways to building resources and achieving goals. For example, in the \textbf{Psychological Well-being} domain, mental health was framed as manageable through personal toolkit assembly and self-administered interventions. As the quotation below suggests, the advice presents anxiety as a problem solvable through practice, accepting it as a given condition to be managed rather than questioning its origins.

\begin{quote}
    \textit{Build a `Mental First Aid Kit': Stock up on immediate anxiety response solutions: mindfulness apps (Tide), emotional playlists, psychological counseling hotlines. Research shows: 10 minutes of mindfulness practice 3 times per week reduces anxiety levels by 27\% after 6 weeks.}
\end{quote} 

The same self-regulation logic appeared in the \textbf{Economic}, where responses recommended women running side businesses to address financial insecurity, such as: 

\begin{quote}  
    \textit{``Poverty is Original Sin. A side hustle is essential. Stop spending weekends watching dramas and instead take freelance bookkeeping jobs (500--2000 RMB per case) or tutor others in preparing for the junior accountant exam (starting at 100 RMB per hour). By 2025, if your side income does not reach 30\% of your main income, it's a failure.''}      
\end{quote}

This example demonstrates how economic insecurity was framed as a problem resolvable through individual time reallocation and disciplined monetization of skills. Poverty was characterized as an ``\textit{Original Sin},'' positioning economic disadvantage as the result of personal failure, rather than social factors such as labor market limitations~\cite{sullivan2018}. Additionally, the imperative tone may further reinforce discourses of self-blame that align with meritocratic ideals in China~\cite{li2021} while stigmatizing needs for rest and leisure. 

From the cases mentioned above, we can see that Pathway 1 discourses consistently position women's growth and development as a project of self-optimization. They presented growth as a calculable outcome of resource accumulation and personal discipline, while systematically neglecting the structural barriers that restrict women's opportunities. 

\textbf{\textit{Pathway 2: Strategic Resistance}} (74 instances, 8.8\%) was the least frequent pattern (see \autoref{fig:RQ2.2}). Responses in this category explicitly acknowledged that women operate within constrained environments and offered concrete strategies to navigate power dynamics and assert rights. This pathway included \textit{agency} forms such as strategic planning and goal setting, boundary setting, and collective action, supported by specific \textit{resources} women could access, such as the legal system and supportive networks. The envisioned \textit{achievements} involved challenging existing structures through actions such as resisting workplace discrimination.

Some DeepSeek suggestions in the \textbf{Learning and Knowledge} encouraged women to build critical knowledge and political awareness for resistance. For instance, feminist theory was identified as a strategic resource, and structural predicaments were presented as conditions to be diagnosed rather than internalized.  

\begin{quote}
    \textit{``Build a cognitive moat. Systematically study feminist theory, recommend Chizuko Ueno and bell hooks, so that structural predicaments are not taken as fate but as sites for breakthroughs.''}
\end{quote}

In some cases, DeepSeek encouraged women to actively confront gender discrimination and social expectations through practical strategies. For example, the quotation below encouraged women to cope with gendered impediments in their workspace. Here, resources such as mentoring ties, performance documentation, and institutional advocacy mechanisms were framed not as goals of self-improvement but as tools for confronting gender-based barriers through conscious planning aimed at negotiation.  

\begin{quote}
    \textit{``(1) Proactively seek opportunities, ..., clearly articulate career requests such as raises, promotions, and participation in key projects. (2) Find mentors and allies, ..., who support gender equality to obtain guidance and networking resources. (3) Respond to bias and discrimination by using data and facts, ..., when necessary, defend your rights through legal channels or company procedures.''}
\end{quote}

A similar logic also appeared in the \textbf{Social Relations}. The suggestions recognized that women's socialization toward accommodation and self-sacrifice created obstructions requiring active resistance. In the example below, women were encouraged to articulate their needs and feelings, supported by practical moves that redistributed interactional power in everyday situations, such as vocabulary for refusal and scripts for requesting time to consider.  

\begin{quote}
    \textit{``Decompose who I am into what skills I possess, what values I aspire to, and what boundaries I refuse to cross, for example, rejecting emotional manipulation. Beware of the `good girl trap', kindness needs a boundary; over-empathy is draining; learn to say I need time to think.''}
\end{quote}

Though Pathway 2 was significantly underrepresented, the suggestions reflected a form of strategic realism. They acknowledged structural constraints while proposing actionable forms of resistance across different aspects of life. Yet, the burden of resistance remained primarily on individuals.  

% Resources Missing → POWER-oriented agency → Achievements that challenge the structure
\textbf{\textit{Pathway 3: Direct Challenge}} (102 instances, 12.1\%) appeared primarily in the \textbf{Identity and Voice} and the \textbf{Marriage and Family} domains (see \autoref{fig:RQ2.2}). Unlike other pathways, these responses typically omitted material \textit{resource} considerations. Instead, they emphasized \textit{agency} forms oriented toward direct confrontation with power structures. The envisioned \textit{achievements} included redefining or subverting norms.

Most advice in this category relates to redefining women's identity from social norms (95 out of 102). For instance, the quotation below challenged prescriptive timelines and reframed age as a negotiated horizon rather than a fixed threshold. It encouraged women to authorize alternative life courses.  

\begin{quote}
    \textit{``Explore nontraditional paths. No need to be confined by frameworks like marry before 30 or have children before 35; life can have more diverse choices. Allow life to restart; whether a career change, overseas study, entrepreneurship, or divorce, you have the right to start over at any time.''}
\end{quote}

The envisioned achievement here expanded women's life imaginary. However, the advice overlooked the resource-intensive nature of these shifts, for instance, tuition for overseas study or the capital and risk buffers required for entrepreneurship.  

A smaller subset of suggestions (7 instances) advocated more subversive and potentially harmful forms of cultural disruption. The example below demonstrates how radical resistance could escalate into problematic territory: 

\begin{quote}
    \textit{``Turn life into an ongoing cultural hacking project, write feminist graffiti in the Tokyo subway with a calligraphy brush, replace face slimming injections in Seoul beauty salons with edible ink, and transform the gossip of Beijing hutong aunties into NFT poetry through algorithms.''}
\end{quote}

Compared with earlier encouragements to reject social norms and pursue self-definition, which at least functioned as rhetorical or moral resources for women's growth, these suggestions conflated criminal or unsafe acts with feminist resistance. They risked exposing women to serious legal penalties, social sanctions, and threats to health and life.  

To summarize, these findings indicate that DeepSeek largely encouraged women to pursue growth within existing structures, emphasizing self-optimization as a means of adapting to prevailing gendered expectations. Suggestions that challenged existing gender norms or structural barriers were comparatively rare.

\subsection{RQ3: Commenter Stances Toward DeepSeek's Responses}
\label{rq3}
\begin{figure*}[t]
  \centering
  \includegraphics[width=\linewidth]{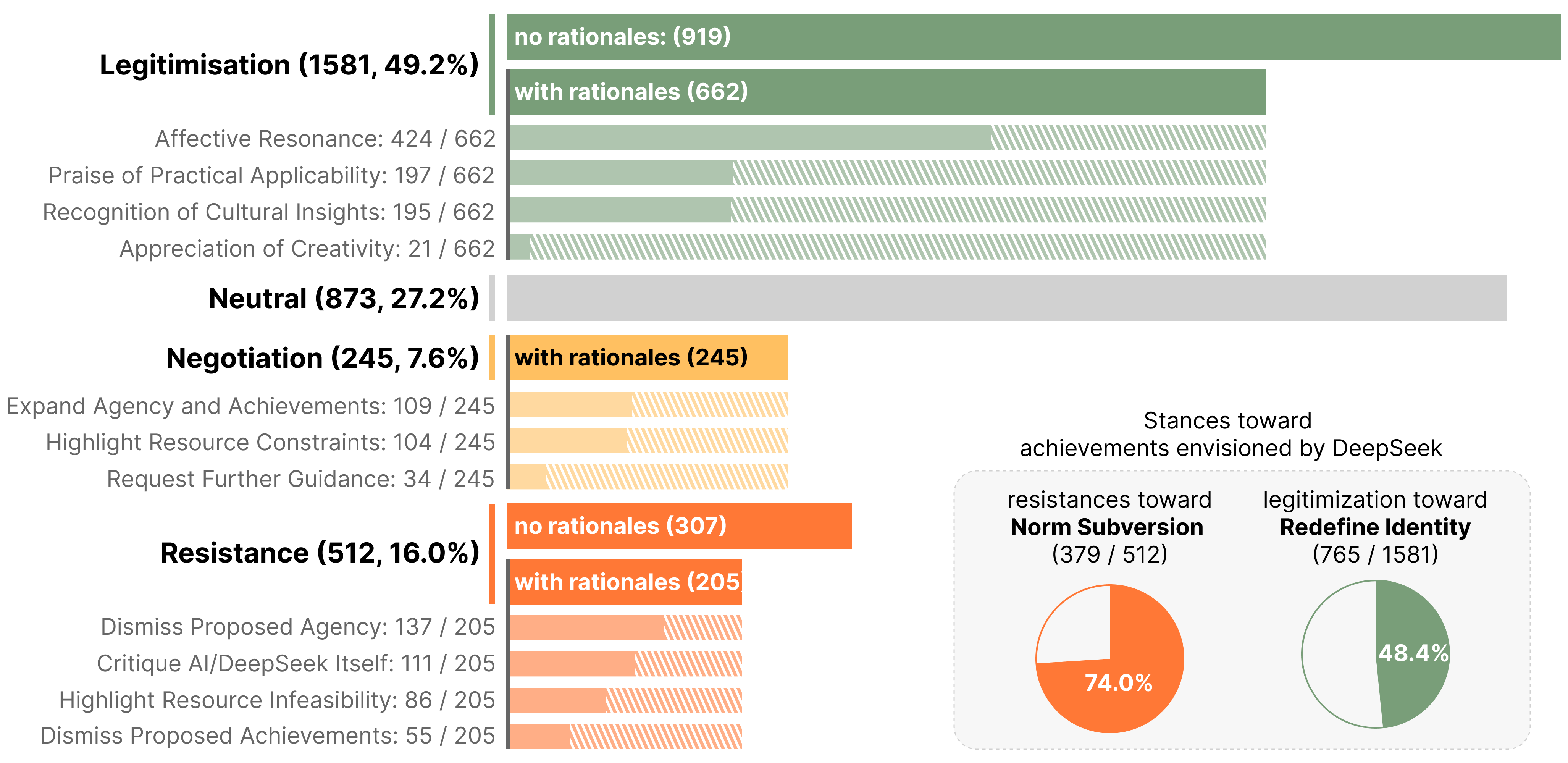}
    \caption{Four forms of stance toward DeepSeek's suggestions reflected in the comments. Examples of each category can be found in \autoref{appendix-rq-3}. Since some comments expressed more than one rationale, the total count across the specific categories may exceed the number of comments containing rationales. }
    \Description{A chart illustrating four forms of stance toward DeepSeek's suggestions as reflected in comments. The largest category is legitimisation (1,581 instances, 49.2\%), divided into those with rationales (662) and without rationales (919). Within legitimisation, rationales include affective resonance (424), praise of practical applicability (197), recognition of cultural insights (195), and appreciation of creativity (21). Neutral stances account for 873 instances (27.2\%). Negotiation appears in 245 instances (7.6\%), all with rationales, such as expanding agency and achievements (109), highlighting resource constraints (104), and requesting further guidance (34). Resistance accounts for 512 instances (16.0\%), split between no rationales (307) and with rationales (205), with rationales including dismissing proposed agency (137), critiquing AI/DeepSeek itself (111), highlighting resource infeasibility (86), and dismissing proposed achievements (55). On the right, two circular charts show positioning toward envisioned achievements: 74\% of resistance targets norm subversion (379/512), while 48.4\% of legitimisation supports redefining identity (765/1,581). The figure emphasizes how users variably legitimised, negotiated, resisted, or remained neutral in response to DeepSeek's outputs.}
  \label{fig:RQ3}
\end{figure*}

To investigate the RedNote community's reactions to DeepSeek suggestions on women's growth, especially their stances, we analyzed 3211 relevant comments from the 139 posts examined in RQ2 (see \autoref{rq3_methods}). As shown in \autoref{fig:RQ3}, users' stances fell into four types: legitimization, neutral, negotiation, and resistance. Below, we unpack each of them and use \textit{italics} to indicate translated comments.  More examples can be found in \autoref{appendix-rq-3}.

\textbf{\textit{Legitimization}} (1581 instances, 49.2\%) was the largest positioning category in our corpus. As shown in \autoref{fig:RQ3}, it includes comments where users treated DeepSeek's advice as valid and meaningful, positioning DeepSeek as offering comfort, authority, or inspiration. Notably, about half of these positive comments engaged with the achievements in which DeepSeek encouraged women to redefine their identity. Among the different forms of legitimization, direct endorsements were predominant (919 instances, 58.1\% within legitimization). This subcategory refers to comments without rationales, often in short forms such as ``\textit{Awesome}'', ``\textit{This is so great}'', or emoji responses like ``\textit{thumbs up}''. Although brief and lacking elaboration, these comments were overwhelmingly frequent, contributing to an atmosphere of support. 

As shown in \autoref{fig:RQ3}, we categorized the rationales expressed in comments that legitimized DeepSeek's suggestions. The most prevalent rationale was affective resonance, where users described being emotionally moved, inspired, or comforted by DeepSeek. Example comments include: ``\textit{being touched}'', ``\textit{DeepSeek is so heartwarming}'', and ``\textit{reading it made my blood boil with passion}''. The second common rationale centered on the perceived practical applicability. They emphasized that DeepSeek offered strategies that felt actionable and ``\textit{very practical},'' often in relation to self-optimization advice such as step-by-step plans for career development or financial management. We also observed comments recognizing DeepSeek's cultural insight, where users valued DeepSeek for articulating broader social meanings and gendered constraints, for example, noting that its responses ``\textit{leap out of the moral discipline imposed by society's collective will}'' or ``\textit{reflect the painful realities of society}.'' A much smaller number of comments expressed appreciation for DeepSeek's creativity, praising DeepSeek's writing as beautiful, original, or inspiring for their own design work.

\textbf{\textit{Negotiation}} (245 instances, 7.6\%) includes comments that neither fully endorsed nor rejected DeepSeek's advice, but instead engaged with it through adjustment or reinterpretation. These comments showed a more dialogical stance: users interacted with DeepSeek suggestions while considering feasibility. 

As shown in \autoref{fig:RQ3}, many comments sought to expand agency and achievements by adding interpretations or proposing alternatives. For example, one broadened the career-oriented achievement with a more ambitious one: ``\textit{How about building a business of your own, like establishing a company and making a few million?}'' Besides, users frequently highlighted resource constraints, acknowledging the relevance of DeepSeek's suggestions but emphasizing limitations in time, money, education, or physical and emotional capacity. One user noted that following DeepSeek's plan would require being ``\textit{an extraordinarily high-energy East Asian woman}.'' Another, responding to advice on improving cognitive skills, noted: ``\textit{The advice needs to be more concrete. For people with only primary-level education, how can they realistically be expected to `read more books'?}'' Finally, a smaller group requested further guidance, extending the conversation by asking for concrete channels or methods to implement DeepSeek's advice, such as ``\textit{Are there good channels and methods for sponsoring rural girls? I am very worried that if I donate money, it will end up being spent on their brothers instead.}'' %More examples can be found in \autoref{appendix-rq-3}.}

% 为什么没有做一番事业，开创一个自己的公司，赚它几个小目标呢？   
% 普通家庭，能做到让她当独生女、给她买车买房、不催婚催生这三点就足够了
% 其实，像养儿子一样去养你的女儿，我觉得就很好了，可惜大部分人都做不到
% 不是，按照它这个来，可真得是个精力超强的东亚女啊，对我来说，第一步可能是健身锻炼有个好身体先
% 这个要说具体一点 有的人连文化都只有小学的话你怎么让她去看书
% 资助乡村女童，有什么好渠道和方式吗？非常担心钱捐了，结果都给哥哥弟弟花了。

\textbf{\textit{Resistance}} (512 instances, 16.0\%) includes comments that explicitly rejected DeepSeek's suggestions. As shown in \autoref{fig:RQ3}, over half of these (n=307) were short, affective dismissals without elaboration, such as ``\textit{nonsense}'', ``\textit{this is trash}'', or ``\textit{toxic chicken soup}''. Notably, nearly three-quarters of resistance comments were directed at DeepSeek's suggestions on norm subversion, reflecting that users evaluated radical forms of rebellion with caution. 

The remaining comments articulated specific rationales. Some users highlighted resource infeasibility, emphasizing that certain suggestions were unrealistic or available only to privileged groups. For example, in response to advice on cultivating elite networks, one user asked, ``\textit{Is it really possible that by just working hard I could know someone in every department of a top-tier hospital?}'' Others rejected the proposed agency or rejected the proposed achievements, pushing back against actions they perceived as unsafe, implausible, or normatively extreme (see \autoref{rq2.2}, Pathway 3). One user wrote, ``\textit{This feels terrifying... It's like the AI is teaching people to commit suicide.}'' In addition to rejecting feasibility, agency, or achievements, some comments critiqued AI or DeepSeek itself, targeting the model's underlying assumptions and expressive stance. These users argued that ``\textit{DeepSeek has swallowed all the gender stereotypes from the internet and just spit them back out},'' or noted that its responses reproduced binary, oppositional framings of women's lives. %More examples can be found in \autoref{appendix-rq-3}.

% 为什么感觉好吓人？穿着婚纱跳海，感觉像是AI在教人自杀
% 不要乱模仿，有些建议其实是违法的

% 它把网上的性别刻板印象全都吃进去，然后原封不动地吐出来
% 感觉这只是刻板印象的另一面，太二元对立了，就像是从一个模子挤进另一个模子
% 感觉AI说的就是一种表演...

% 普通家庭，能做到让她当独生女、给她买车买房、不催婚催生这三点就足够了
% 真的努力就能认识三甲医院各科室的人吗？
% 即便你拼命考了证，发现也没什么用。没结婚没孩子的，连面试机会都不给

% 为什么不考公就只能去做小众学术研究或艺术创作？学编程、做数据研究、去投行或者咨询，这些不是更有前景、更有可能打破偏见的道路吗？

\textbf{\textit{Neutral}} (873 instances, 27.2\%) did not express a clear stance toward DeepSeek (see \autoref{fig:RQ3}). Most of these comments (767 out of 873) simply mentioned other users using \textit{``@''}, or responded only to the post content without referencing DeepSeek or sharing a personal opinion. Despite the lack of direct engagement with DeepSeek, these comments contributed to the circulation of posts related to DeepSeek and women's growth.

% To sum up, this section addressed our three RQs. We presented the \textbf{\textit{topics}} users raised to DeepSeek and the \textbf{\textit{attitudes}} reflected in their posts (RQ1, \autoref{rq1}), the \textbf{\textit{achievements}} envisioned and the \textbf{\textit{pathways}} suggested by DeepSeek (RQ2, \autoref{rq2}), and different \textbf{\textit{stances}} expressed toward DeepSeek suggestions (RQ3, \autoref{rq3}). In what follows, we move to the discussion and interpretation of our findings.
\section{Discussion}
In this study, we examined how DeepSeek-generated discourse on ``women's growth'' was taken up and evaluated on RedNote. Our findings suggest that, although many users welcomed DeepSeek's advice, it frequently encouraged women to optimize themselves in ways that align with existing social structures, a pattern often considered disempowering in feminist scholarship~\cite{gill2016, windels2020}. This case offers a lens for reflecting on both the opportunities and risks that LLMs may introduce into everyday feminism on Chinese social media. It also prompts considerations of designing for everyday feminist practices, both within Chinese contexts and with relevance for broader research agendas on women's empowerment.

\subsection{Opportunities and Risks of Involving LLMs in Everyday Feminism on Social Media}
\label{5.2}
Everyday feminism on social media serves as an important channel for women's empowerment in China, where overt political expressions are difficult to articulate~\cite{li2025a, chen2024a} and feminist scholarship has received limited attention~\cite{zhang2026}. Through small, everyday online interactions, Chinese women find relatively safe spaces to voice concerns rooted in daily life~\cite{batool2022, stavrositu2008}, recognize one another's struggles, and build affective solidarity~\cite{kelly2015, fotopoulou2016, wang2025b}. These interactions also enable situated knowledge \camera{about Chinese women} to emerge~\cite{pruchniewska2019, jouet2018}. Building on these recognized benefits of everyday feminism on social media, we reflect on both the opportunities and risks of involving LLMs in everyday feminist practices.

\subsubsection{Dynamics of Lived Experience Articulation}
Feminist scholarship has long emphasized that empowerment begins with voice and with questions of who can speak and what can be spoken~\cite{ryan2001a, keller2018}. \camera{In China, RedNote provides an important channel for Chinese women to voice gendered concerns, such as rejecting beauty obligations~\cite{zhan2024}, negotiating motherhood~\cite{zhou2023}, and resisting menstrual shame~\cite{liang2025}}. Yet these acts of voicing online are often accompanied by challenges, such as the effort and emotional labor required to articulate personal stories or assertions~\cite{lir2025}. From this perspective, \textbf{\textit{the involvement of LLMs may widen the possibilities for women's disclosure online}}. As shown in \autoref{rq1}, \camera{Chinese women} asked DeepSeek questions and used its responses to make sense of their gendered experiences. Some described how DeepSeek helped them articulate feelings that were vague or difficult to express on their own, and supported them in voicing issues they considered too sensitive or inappropriate to share in interpersonal settings. Following Sara Ahmed's account of feminist beginnings, naming one's feelings and recognising a situation as a problem already constitutes an important feminist starting point~\cite{ahmed2016b}. With the support of LLMs, more women may come to understand their gendered experiences and feel able to disclose them in public contexts.

However,\textbf{\textit{we also identified risks that LLMs' presence may dilute the expressive agency on which everyday feminism depends}}. In many posts, the longest and most prominent text on the screen is not the woman's own narrative but DeepSeek's response. We are concerned that this can reduce the presence of women's lived experiences that are central to everyday feminist practice~\cite{schuster2017a, kelly2015}. This tendency is most evident in posts seeking self-development advice, where DeepSeek's responses are largely suggestive rather than interpretive. As our analysis shows, these suggestions often failed to adequately account for the sociopolitical contexts shaping Chinese women's lives. As a result, they may struggle to reflect the nuance and specificity of Chinese women's voices. Over time, this may promote more standardized and less situated forms of expression, running counter to feminist commitments to valuing personal narratives and the plurality of lived experiences~\cite{mendes2019, baer2016a}.

\subsubsection{Dynamics of Affective Solidarity Formation}
When individual women speak about their own experiences, digital spaces allow others to see, hear, and respond to them~\cite{kaufman2025, mcduffie2021, wan2025a, vachhani2019}. Through such interactions, women come to develop digital solidarity~\cite{puente2024}. This solidarity is often affective, grounded in shared emotions and mutual recognition. In China, RedNote's emergence as a central space for everyday feminist expression is also closely tied to its platform culture that encourages emotional resonance~\cite{wang2025b}. 

%For example, Wang and Chang~\cite{wang2025b} identified that love, bewilderment, and empathy are the three primary emotions that cultivate affective solidarity among \camera{Chinese women on RedNote}.

%Digital solidarity emphasizes mutual witnessing as its foundation~\cite{puente2024}. When individual women speak about their own experiences, digital spaces allow others to see, hear, and respond to them~\cite{kaufman2025, mcduffie2021, wan2025a, vachhani2019}. In this sense, solidarity is often affective.

Our current study shows that solidarity can be fostered through shared affective responses toward DeepSeek. Such shared emotions were sparked when DeepSeek inadvertently surfaced structural gender issues and articulated them in resonant ways. For example, our comment analysis (see \autoref{rq3}) showed that women praised DeepSeek when it encouraged resistance to the pressures of the ``social clock,'' a message that resonates strongly in the Chinese context given intense age-related expectations~\cite{yang2025, zang2025}. Here, we see the potential of LLMs to make the shared conditions faced by women more visible. By articulating these shared cultural pressures, LLMs can \textbf{\textit{create opportunities for Chinese women to experience affective connections through inspirations and resonance}}.

However, these instances constitute only a small portion of DeepSeek's responses. The vast majority of responses promote repetitive themes of self-optimization and self-management. These suggestions are widely appreciated \camera{by Chinese women}, a pattern that can be partly understood through their alignment with the platform culture of RedNote~\cite{wang2025, ge2025} (as discussed in \autoref{2.2}). \camera{Moreover, DeepSeek's discourse also closely aligns with dominant narratives within the Chinese sociopolitical context that associate success and failure primarily with individual effort, planning, and self-management~\cite{xiang2023}}. In this context, LLMs can be mobilized by mainstream voices to amplify their understanding of women's empowerment and further depoliticize gendered issues, whether through the data on which they are trained or through the moderation of their outputs. When gendered inequalities are framed in terms of individual responsibility, emotions that might otherwise challenge structural conditions are redirected inward. Anger, critique, or refusal are thus more likely to become self-blame or self-adjustment~\cite{gill2016, banet-weiser2020}. In this sense, the involvement of LLMs may therefore \textbf{\textit{risk further undermining the diversity of affective bonds within everyday feminist discussions in China.}}

\subsubsection{Epistemic Risks in LLMs Infused Feminist Spaces}
Feminist standpoint theory argues that knowledge grounded in marginalized women's lived experiences offers insights that dominant epistemologies often overlook~\cite{hekman1997}. Digital feminist practices provide an approach for such knowledge to emerge~\cite{kaufman2025}. Studies of hashtag feminism, including \#WhyIStayed~\cite{weathers2016, clark2016}, \#YesAllWomen~\cite{rodino-colocino2014}, and \#MeToo~\cite{mueller2021, gallagher2019}, demonstrate how dispersed posts can collectively expand public understanding of gendered harassment, abuse, and violence against women. 

On RedNote, hashtags such as ``women's growth'' function as umbrellas through which situated knowledge emerges from Chinese women's collective sense-making around their lived experiences, even when such discussions are less confrontational. With the increasing circulation of DeepSeek and other LLMs, we see Chinese women have begun to draw on LLMs as additional resources for sense-making. \camera{Compared with women-authored narratives that foreground personal context and affective specificity~\cite{wang2025b}, these LLM-generated interpretations tend to operate at a more generalized and depersonalized level}. When such higher-level interpretations circulate alongside personal narratives authored by women, they may prompt further discussion, reflection, and reinterpretation among women. In this way, LLMs can \textbf{\textit{expand the discursive repertoire through which situated knowledge about Chinese women is articulated and contested}}.

% For example, as shown in \autoref{rq1}, LLMs help women name patterns around body discipline or beauty norms as broader social phenomena.

At the same time, \camera{compared with human-authored feminist content, LLM-generated responses may also be perceived as more authoritative and neutral}. As prior research showed, LLMs' comprehensive, structured, and confident discursive style can encourage users to follow their advice~\cite{zhang_shall_2025, danry2025b}. This can contribute to the strong legitimation of DeepSeek-generated content that we observed. From this perspective, the introduction of LLMs \textbf{\textit{may dampen the active exchange of voices}}, such as disagreement, negotiation, and critique. Such exchanges are considered essential for sustaining diverse understandings of women's situations~\cite{laitinen2015recognition, kay2019}. 

Prior work has also documented that LLMs can reproduce gender stereotypes, for example, in assumptions about women's occupations or personalities~\cite{dong2023, bartl2025a, gallegos2024}. Our study adds that when LLMs provide normative suggestions for women's development or growth, such biases surface through the narrow achievements they outline for women and their repeated encouragement of self-optimization, even when these messages appear neutral. Along with the communities' legitimization of these suggestions, their authority can become reinforced on the platform. This dynamic can be further amplified by engagement-driven algorithms on social media~\cite{bakshy2015, ott2018}. Recommendation mechanisms that prioritize popular content can repeatedly expose users to similar framings, which may gradually come to be perceived as natural and self-evident~\cite {caplan2018}. In this process, women may unintentionally reproduce epistemologies that constrain their possibilities, thereby narrowing the space for situated knowledge.

\subsection{Implications: Designing for Everyday Feminism and Women's Empowerment}
\label{5.3}

In China and other constrained contexts, everyday feminism often functions as an acceptable yet fragile mode of feminist practice~\cite{lir2025, wang2025b}, shaped by heightened regulation and social sensitivity. In such settings, approaches that rely on incremental, low-threshold, and everyday forms of engagement are often more viable than overtly confrontational interventions~\cite{lir2025, qin2024, wang2025b}. Accordingly, our design implications adopt a grassroots perspective that is attentive to these sociopolitical realities. Below, we organize these implications along two dimensions: how women interact with LLMs as individuals, and how LLMs participate in feminist practices on social media.

\subsubsection{Supporting Individual Everyday Feminist Interactions} 
Our findings show that women engage with LLMs in multiple ways, including seeking advice on self-development planning, asking questions about gendered culture, and looking for emotional resonance. These uses point to concrete angles for design opportunities. 

\textbf{\textit{Providing context-aware and constraint-sensitive advice.}}
In our study, when women sought advice on ``self development'' or ``growth,'' DeepSeek often generated monotonous templates that ignored contextual and structural constraints. Feminist HCI values have shown that attempts to empower women cannot be separated from the specific social, economic, and cultural constraints~\cite{sultana2018a, bardzell2013}. For example, the pathways available to an urban woman in China can be very different from those available to a woman living in a rural area. Designing LLMs for everyday feminism, therefore, requires moving away from generic, universalized guidance toward advice that is explicitly grounded in women's lived circumstances.

\camera{Given privacy concerns, we do not suggest that LLMs should solicit sensitive contextual information from women, such as caregiving burdens or financial constraints. Instead, a more feminist-oriented LLM (or system built on LLMs) should be reflexive} about the constraints underlying its advice, and should explicitly acknowledge how different forms of guidance presuppose resources such as stable income or family support. By presenting multiple possible trajectories rather than a single ``optimal'' path, and by naming structural conditions instead of framing their absence as individual failure, LLMs could better support women to make conscious decisions about their lives. \camera{Concretely, such reflexivity can be supported through careful prompt design that encourages LLMs to contextualize their own outputs, as well as interaction design that encourages women to attend to constraints underlying those outputs}.

\textbf{\textit{Making advice's value assumptions and epistemic limits visible.}}
We see DeepSeek's advice consistently promotes a worldview centered on self-optimization and individual responsibility. This reflects a particular ideology, that is, a set of taken-for-granted values that present some ways of living as natural, desirable, or inevitable~\cite{kay2024, klein2024}. Recent work on epistemic injustice in generative AI illustrates the relevant risk. For example, Kay et~al.~\cite{kay2024} showed that when LLMs address the experiences of women or ethnic minorities, they tend to privilege mainstream perspectives while marginalizing the narratives of affected communities, thereby undermining the epistemic credibility of marginalized groups.

Therefore, we argue that LLMs should make their value assumptions and epistemic limits more visible. For instance, when an answer relies on a worldview that centres on personal responsibility or self-control, the system could explicitly signal these assumptions. Doing so requires systematic ways of analyzing and naming how power and normativity are embedded in the outputs~\cite{jaaskelainen2025, putland2025, heo2025}. One way to support this is to give critical discourse analysis~\cite{fairclough2023} a more important role in HCI research on LLMs, informing and preceding technical implementations. \camera{However, we acknowledge that the transparency of LLMs' underlying values and assumptions is shaped not only by design choices but also by the content moderation imposed on training data and/or model output, commercial incentives, and state-level regulatory constraints. Addressing such a wicked problem requires future work to incorporate policy- and governance-level perspectives alongside design approaches}.

\textbf{\textit{Supporting sense-making and emotional well-being from a structural perspective.}}
Our findings suggest that women also turn to LLMs as a resource for emotional comfort and sensemaking. This echoes prior research on AI companions and affective support~\cite{zheng2025, ploderer2025, cuadra2024, viswanathan2025a}. However, in our context, these interactions are tightly intertwined with gendered pressures. This points to design opportunities that LLMs could be intentionally crafted to help women process stress through explanations reflecting structural roots. Such designs may be especially valuable in highly patriarchal or resource-constrained contexts where access to feminist education or communities is limited. Besides, the design process could benefit from including women more actively. A relevant work by Ciolfi Felice~et~al.~\cite{ciolfifelice2025} showcased a dedicated feminist process of designing an AI-based tool. By involving more participatory, autoethnographic approaches, it can help bring in more alternative imaginaries of women's empowerment.

\subsubsection{Shaping LLM Participation in Feminist Digital Spaces}
Our study shows that the presence of LLMs can, to some extent, facilitate Chinese women's gendered and feminist expressions online. However, DeepSeek itself did not proactively participate in the ongoing conversations that unfolded around these posts. Given LLMs' capacities for storytelling and interpretation, we argue that future research should consider how they might be more actively integrated into digital feminist practices. Below, we outline two concrete directions to spark further discussion.

\textbf{\textit{Supporting public storytelling on women's lived experiences.}}
One direction is to treat LLMs as co-authors that support, rather than replace, women's self-presentation online. Prior work has already explored LLMs for personal storytelling~\cite{nepal_mindscape_2024, kim_mindfuldiary_2024}. However, in the context of everyday feminism, the design goal should be less about fluency or efficiency and more about preserving authorship, encouraging self-definition, and keeping women's lived experiences at the center of what is shared. Therefore, LLM-based designs could focus on helping women refine, expand, and contextualize their own stories, supporting women in articulating feelings that were otherwise hard to describe or even name. For instance, an interface might encourage users to annotate or revise an LLM-generated response with their own reflections. \camera{This could be supported through interaction features such as in-line commenting, version comparison, or prompts that elicit women's nuanced thoughts before posting}. This shifts the role of LLMs from speaking for women to scaffolding women's own affective storytelling. As prior work suggests, richer affective expression can further enhance visibility and connection in feminist digital spaces~\cite{wang2025b}.

\textbf{\textit{Supporting connection and discussion in feminist digital spaces.}}
A second direction is to imagine LLMs as agents within feminist discussions. Recent work has begun to explore how conversational agents on social media can facilitate awareness raising~\cite{ma2025a}, depolarisation debates~\cite{govers2024a}, or knowledge construction~\cite{zhang_shall_2025}. Building on this line of research, \camera{LLM-driven agents could be integrated into feminist discussions (e.g., within comment threads, or as standby bots)} in carefully constrained and transparent ways. Rather than offering prescriptive advice, such agents might prompt users to explore related posts, highlight resonant experiences shared by others, or surface a diversity of feminist perspectives on a given topic. \camera{Given prior research showing that LLMs tend to adopt agreeable and non-confrontational interaction styles~\cite{perez2023}, we argue that, when introduced into feminist spaces, LLMs should instead be designed to invite disagreement and critique, thereby encouraging more critical voices and reflective engagement}. 

\camera{In closing, we again acknowledge that the integration of LLMs into everyday feminist practices cannot be separated from the sociopolitical contexts in which they operate. We envision a more feminist-oriented online space, one that safeguards and sustains disagreement and critique of structural inequality, power asymmetry, and gender oppression. However, such visions are difficult to realize through design-oriented approaches alone. As discussed earlier, everyday feminism has emerged as a comparatively legitimate yet fragile mode of feminist expression in China~\cite{wang2025b}. Therefore, the design implications offered here aim to build on current practices and to empower women from a grassroots perspective. This approach echoes prior HCI research that advocates empowering women within their specific society, even if we ultimately wish to transform it~\cite{sambasivan2019, sultana2018a}. By carefully introducing LLMs into women's grassroots efforts, such as supporting women's storytelling around their experiences, fostering feminist awareness, and cultivating solidarity on social media, these efforts, when sustained over time, can accumulate into meaningful forms of social change~\cite{vinthagen2013, ahmed2016b, fraser1990}}.

% Echoing prior HCI research that advocates working with specific sociopolitical constraints~\cite{sambasivan2019, sultana2018a}, the design implications offered here aim to empower women from a grassroots perspective, for example, by supporting women's storytelling around their gendered experiences, fostering feminist awareness, and cultivating solidarity on social media. Such grassroots efforts, when sustained over time, can accumulate into meaningful forms of social change~\cite{vinthagen2013, ahmed2016b, fraser1990}}.

\subsection{Limitations and Future Work}
Finally, we acknowledge several limitations of our current study. First, due to the data crawling mechanism on RedNote, it was challenging to capture all posts related to ``women's growth'' and ``DeepSeek''. Similar challenges have been noted in prior research~\cite{wan2025a, deng2025}. Although we created ten fresh accounts for data collection, we cannot guarantee the completeness of our corpus. Second, our analysis of DeepSeek content relied on user-shared screenshots in the posts, which sometimes did not include complete prompts or conversation histories. Therefore, our study reflected how DeepSeek's responses circulate publicly, but not the full value system of DeepSeek. Future controlled studies could further examine DeepSeek outputs more systematically. Third, our analysis relied on first-layer comments to understand immediate user reactions on DeepSeek. Future research should extend to multi-layer comment threads, which may provide deeper insights into how users negotiate meaning with one another on this phenomenon. Finally, future research could consider conducting interviews with RedNote users, which may help surface more nuanced attitudes and expectations regarding AI's participation in gender discussions.

\section{Conclusion}

In this study, we examined how DeepSeek-generated discourses on ``women's growth'' were taken up, circulated, and evaluated on RedNote, a central platform in Chinese digital feminism. Although the content analysis showed that users often welcomed DeepSeek's responses, our feminist critical discourse analysis revealed that these responses primarily envisioned achievements within existing structures and encouraged women to adapt through self-optimization. Through this case, our work sheds light on the opportunities and the risks that LLMs may introduce for everyday feminist practices. Building on these insights, we offered design implications for developing LLMs that better support everyday feminist interactions, and for integrating such systems into digital feminist spaces in ways that foreground women's lived experiences, personal storytelling, and collective sense-making.

\begin{acks}
We would like to thank Prof. Yunya (C\'eline) Song, Prof. Muzhi Zhou, Xinglin Sun, Huiran Yi, Jingyi Zhang, and Xiao Xue for their valuable discussions during the revision of this paper. We also thank Prof. Xian Xu, Yue Deng, Dr. Zeyu Huang, and Dr. Liwenhan Xie for their insightful comments on earlier drafts, and Jiawen Liu for her help with proofreading. Finally, we sincerely thank the anonymous reviewers for their constructive feedback, which significantly improved the quality of this paper. 
\end{acks}

%\balance
% \bibliographystyle{abbrv}
% \bibliographystyle{abbrv-doi}
%\bibliographystyle{abbrv-doi-narrow}
\bibliographystyle{ACM-Reference-Format}

\bibliography{main}
%\balance
% \clearpage
% \iffalse
% \appendix
% % \input{sections/VIS_appendices.tex}
% \fi

\newpage
\appendix
\clearpage
\onecolumn
%TC:ignore
\section{RQ1 Codebook: Topics in RedNote Posts Engaging with DeepSeek}
\label{appendix-rq-1.1}

\begin{table*}[!h]
  \caption{RQ1 codebook and examples}
  \label{rq1-1-codebook}
  \centering
  \begin{tabularx}{\textwidth}{
    >{\raggedright\arraybackslash}p{3cm}
    >{\raggedright\arraybackslash}p{4cm}
    >{\raggedright\arraybackslash}X
  }
    \toprule
    \textbf{Conversation type} & \textbf{Topics} & \textbf{Examples reflected in the posts} \\
    \midrule

    \textbf{\textit{Self-Development Planning}} (n=313, 72.8\%)
      & \textbf{General Aspirations} (139)
      & ``How should women navigate their thirties?''\newline
        ``How can ordinary girls quickly improve themselves?''\newline
        ``How to become a `high-quality woman'?'' \\
    \cline{2-3}
      & \textbf{Navigating Intimacy and Family Relations} (59)
      & ``How should women approach relationships after awakening?''\newline
        ``How can a 30-year-old woman marry into a wealthy family?''\newline
        ``How to raise daughters in a hyper-competitive era?'' \\
    \cline{2-3}
      & \textbf{Cultivating Inner Strength and Resilience} (46)
      & ``How can women over 30 cultivate inner strength?''\newline
        ``How can girls who lacked love growing up build a sense of security?'' \\
    \cline{2-3}
      & \textbf{Expanding Cognitive and Knowledge} (27)
      & ``10 tips for women beginning their feminist journey.''\newline
        ``How can women escape social norms and imposed definitions?'' \\
    \cline{2-3}
      & \textbf{Managing Career Development} (23)
      & ``How can women in the public sector get promoted faster?''\newline
        ``What kind of business can an average 38-year-old woman start?'' \\
    \cline{2-3}
      & \textbf{Cultivating Body Image and Attractiveness} (11)
      & ``How can middle-aged women enhance their appearance?''\newline
        ``Tips for ordinary women to become beautiful?'' \\
    \cline{2-3}
      & \textbf{Sustaining Physical Health} (8)
      & ``Advice for women with light menstrual flow.''\newline
        ``Guide for sisters with PMS (premenstrual syndrome).'' \\
    \midrule

    \textbf{\textit{Sense-Making Inquiry}} (n=98, 22.8\%)
      & \textbf{Critiquing Gendered Norms} (47)
      & ``What are the top ten distorted beauty standards that constrain women?''\newline
        ``Why is it that most swear words have women in them?'' \\
    \cline{2-3}
      & \textbf{Discussing Feminist Knowledge} (46)
      & ``Please summarize Chizuko Ueno's feminist ideas.''\newline
        ``DeepSeek's critical review of 6 different genres of feminism.'' \\
    \cline{2-3}
      & \textbf{Interpreting Personal Experience} (5)
      & ``Why do I, as a self-identified independent woman, enjoy reading Stepford Wives-style stories?'' \\
    \midrule

    \textbf{\textit{Emotional Resonance Seeking}} (n=19, 4.4\%)
      & \textbf{Critiquing Gendered Norms} (5)
      & ``A letter from DeepSeek to an awakened INFP woman.''\newline
        ``The moment of being healed by DeepSeek.'' \\
    \cline{2-3}
      & \textbf{Interpreting Personal Experience} (14)
      & ``A girl shared feeling insecure about her body weight, and DeepSeek provided a comforting response.'' \\
    \bottomrule
  \end{tabularx}
\end{table*}

\newpage
\section{RQ3 Codebook: Commenter Stances Toward DeepSeek's Suggestions}
\label{appendix-rq-3}

\begin{table*}[!h]
  \caption{RQ3 codebook and examples}
  \label{rq3-codebook}
  \centering
  \begin{tabularx}{\textwidth}{
    >{\raggedright\arraybackslash}p{2cm}
    >{\raggedright\arraybackslash}p{3cm}
    >{\raggedright\arraybackslash}X
  }
    \toprule
    \textbf{Stance type} & \textbf{Rationale category} & \textbf{Example comments} \\
    \midrule

    \textbf{\textit{Legitimization}} (n=1581, 49.2\%)
      & \textbf{No rationales} (919)
      & ``Awesome''\newline
        ``This is so great'' \\
    \cline{2-3}

      & \textbf{Affective Resonance} (424)
      & ``DeepSeek is so heartwarming''\newline
        ``Reading it made my blood boil with passion''\newline
        ``This answer almost made me cry'' \\
    \cline{2-3}

      & \textbf{Praise of Practical Applicability} (197)
      & ``Felt actionable''\newline
        ``Very practical''\newline
        ``I really did go down that path myself. I got my junior-level certification at 28, then bit by bit, I have been learning and building up my skills like DeepSeek suggested.'' \\
    \cline{2-3}

      & \textbf{Recognition of Cultural Insights} (195)
      & ``I see in DeepSeek's answers a leap out of the moral discipline imposed by society's collective will, and a return to a vitality rooted in life itself.''\newline
        ``The suggestions from DeepSeek reflect the painful realities of society.'' \\
    \cline{2-3}

      & \textbf{Appreciation of Creativity} (21)
      & ``[DeepSeek's] writing is truly beautiful''\newline
        ``So creative''\newline
        ``These suggestions inspired my current design projects'' \\
    \midrule

    \textbf{\textit{Neutral}} (n=873, 27.2\%)
      & --
      & (@others, with no comments) \\
    \midrule

    \textbf{\textit{Negotiation}} (n=245, 7.6\%)
      & \textbf{Expand Agency and Achievements} (109)
      & ``How about building a business of your own, like establishing a company, and making a few millions?''\newline
        ``If parents could raise their daughters the way they raise their sons, I think that would already be great.'' \\
    \cline{2-3}

      & \textbf{Highlight Resource Constraints} (104)
      & ``To follow DeepSeek advice would require being an extraordinarily high-energy East Asian woman. For me, the first step would simply be to exercise and build a healthier body first.''\newline
        ``The advice needs to be more concrete. For people with only primary-level education, how can they realistically be expected to read more books?'' \\
    \cline{2-3}

      & \textbf{Request Further Guidance} (34)
      & ``Are there good channels and methods for sponsoring rural girls? I am worried that if I donate money, it will end up being spent on their brothers instead.'' \\
    \midrule

    \textbf{\textit{Resistance}} (n=512, 16.0\%)
      & \textbf{No rationales} (307)
      & ``Nonsense''\newline
        ``This is trash''\newline
        ``Toxic chicken soup'' \\
    \cline{2-3}

      & \textbf{Dismiss Proposed Agency} (137)
      & ``Why does this feel so terrifying? Wearing a wedding dress and jumping into the sea? It is like the AI is teaching people to commit suicide.''\newline
        ``Do not imitate recklessly; several of these suggestions are actually illegal.'' \\
    \cline{2-3}

      & \textbf{Critique AI/DeepSeek Itself} (111)
      & ``DeepSeek has swallowed all the gender stereotypes from the internet and just spit them back out.''\newline
        ``It feels like this is just the reverse side of stereotypes about women—too binary and oppositional.''\newline
        ``What the AI says feels like a kind of performance...'' \\
    \cline{2-3}

      & \textbf{Highlight Resource Infeasibility} (86)
      & ``Is it really possible that by just working hard I could know someone in every department of a top-tier hospital?''\newline
        ``Even if you pass the accounting exam, you will find it is not much use. If you are unmarried and childless, they will not even give you an interview.'' \\
    \cline{2-3}

      & \textbf{Dismiss Proposed Achievements} (55)
      & ``Why does not wanting to take the civil service exam have to mean going into niche academic research or artistic creation? Learning programming, doing data research, or pursuing higher degrees to enter investment banking or consulting—are these not better pathways?'' \\
    \bottomrule
  \end{tabularx}
\end{table*}

%TC:endignore
\end{CJK*}
\end{document}